\documentclass[useAMS, usenatbib]{mnras}
\usepackage{graphicx,amsmath,color,amssymb}

\usepackage[pdftitle={}]{hyperref}

\newcommand{\beq}{\begin{equation}}
\newcommand{\eeq}{\end{equation}}
\newcommand{\barr}{\begin{eqnarray}}
\newcommand{\earr}{\end{eqnarray}}

\newcommand{\Lya}{Lyman-$\alpha$}

\newcommand{\mfemu}{\texttt{MFEmulator}}

\newcommand{\Data}{\mathcal{D}}
\newcommand{\Ssubset}{\mathcal{S}}

\newcommand{\gp}{\textsc{gp}}

\newcommand{\realspace}{\mathbb{R}}

\newcommand{\GP}{\mathcal{GP}}

\newcommand{\normal}{\mathcal{N}}

\newcommand{\dd}{\textrm{d}}
\newcommand{\uniform}{\mathcal{U}}

\newcommand{\kms}{\,\textrm{km\,s}^{-1}}

\newcommand{\mpgadget}{\textsc{mp-gadget}}

\newcommand{\xvec}{\boldsymbol{x}}
\newcommand{\yvec}{\boldsymbol{y}}

\newcommand{\thetavec}{\boldsymbol{\theta}}

\newcommand{\Kvec}{\boldsymbol{\mathrm{K}}}
\newcommand{\kvec}{\boldsymbol{k}}
\newcommand{\Mpc}{\mathrm{Mpc}}
\newcommand{\ard}{\textsc{ard}}

\newcommand{\inputparam}{\{  h, \Omega_0, \Omega_b, A_s, n_s\}}
\newcommand{\gadget}{\textsc{gadget3}}
\newcommand{\mpi}{\textsc{mpi}}
\newcommand{\Mpch}{\,\textrm{Mpc/h}}
\newcommand{\hMpc}{\,h\textrm{Mpc}{^{-1}}}

\newcommand{\lowres}{\textsc{lr}}
\newcommand{\highres}{\textsc{hr}}
\newcommand{\class}{\textsc{class}}
\newcommand{\tacc}{\textsc{tacc}}
\newcommand{\hpcc}{\textsc{hpcc}}
\newcommand{\rvec}{\boldsymbol{r}}
\newcommand{\cov}{\mathrm{cov}}

\newcommand{\npart}{N_\mathrm{ptl,side}}
\newcommand{\boxsize}{L_\mathrm{box}}

\newcommand{\kbin}{k_{j}}

\newcommand{\xyemulator}[2]{#1\,{\lowres}-#2\,{\highres} emulator}

\begin{document}

\title[MF Emulator]{
Multi-Fidelity Emulation for the Matter Power Spectrum using Gaussian Processes}
\author[ M.-F. Ho et al.]{Ming-Feng Ho,$^1$\thanks{E-mail: mho026@ucr.edu} Simeon Bird,$^1$ Christian R. Shelton.$^2$\\
$^1$Department of Physics \& Astronomy, University of California, Riverside,
900 University Ave., Riverside, CA 92521, USA\\
$^2$Department of Computer Science \& Engineering,University of California, Riverside, 900 University Ave., Riverside, CA 92521, USA\\
}

% datetime adjustment
\date{\today}

\pagerange{\pageref{firstpage}--\pageref{lastpage}} \pubyear{2019}
\pagenumbering{arabic}
\label{firstpage}

\maketitle

\begin{abstract}
% Abstract
We present methods for emulating the matter power spectrum by combining information from cosmological $N$-body simulations at different resolutions.
%Estimating cosmological parameters is one of the major purposes of cosmological simulations.
An emulator allows
estimation of simulation output by interpolating across the parameter space of a limited number of simulations.
% us to use a limited number of simulations to estimate parameters by interpolating across the parameter space.
%However, the accuracy of an emulator is limited by the number of expensive high-resolution simulations available.
We present the first implementation in cosmology of multi-fidelity emulation, where many low-resolution simulations are combined with a few high-resolution simulations to achieve an increased emulation accuracy. The power spectrum's dependence on cosmology is learned from the low-resolution simulations, which are in turn calibrated using high-resolution simulations.
%We present a multi-fidelity emulator for the matter power spectrum by constructing the covariance kernels between the Gaussian process emulators from different fidelities.
%Multi-fidelity emulation arises when cheap but biased data are more accessible than expensive but accurate data.
%Thus, a multi-fidelity emulator uses cheap data to have decent parameter coverage while using a few high-fidelity data to calibrate the biased outputs.
We show that our multi-fidelity emulator predicts high-fidelity counterparts to percent-level relative accuracy when using only $3$ high-fidelity simulations and outperforms a single-fidelity emulator that uses $11$ simulations, although we do not attempt to produce a converged emulator with high absolute accuracy.
With a fixed number of high-fidelity training simulations,
we show that our multi-fidelity emulator is $\simeq 100$ times better than a single-fidelity emulator at $k \leq 2 \hMpc$, and $\simeq 20$ times better at $3 \leq k < 6.4 \hMpc$. Multi-fidelity emulation is fast to train, using only a simple modification to standard Gaussian processes.
Our proposed emulator shows a new way to predict non-linear scales by fusing simulations from different fidelities.
\end{abstract}

% keywords --

% article itself
\section{Introduction}

Current and next generation large scale structure surveys, such as
\textsc{des}\footnote{\url{https://www.darkenergysurvey.org}} \citep{DES:2020},
\textsc{lsst} (Rubin Observatory)\footnote{\url{https://www.lsst.org}} \citep{LSST:2002}, \textsc{euclid}\footnote{\url{https://sci.esa.int/web/euclid}} \citep{Euclid:2018}, \textsc{desi}\footnote{\url{https://www.desi.lbl.gov}} \citep{DESI:2016}, and the Roman Space Telescope (\textsc{wfirst}) \citep{Spergel:2013} will probe gravitational clustering and galaxy formation at small scales with high accuracy.
Thus, the future of cosmology relies on exploiting the information in non-linear structure formation at small scales, where numerical $N$-body simulations must be used to give accurate theoretical predictions.

% [Future] we need accurate prediction from N-body due to we want
% percent-level accuracy on the cluster measurements like
% Galaxy clusters or weak lensing, so we can infer DM and DE
Cosmological linear perturbation theory provides accurate analytic predictions on the clustering of mass up to $k \sim 0.1 \hMpc$. % mysteries: nature of dark matter, dark energy, inflation,
% The theory of linear perturbations from primordial fluctuations well describe the current astronomical observations on large scales \citep{Planck:2015}.
Despite the success of the standard model of cosmology,
several fundamental physics puzzles are still unanswered:
the accelerated expansion of the Universe \citep{Caldwell:2009},
the nature of dark matter \citep{Feng:2010},
% the physics of Inflation \citep{Allahverdi:2010},
and the sum of the neutrino masses \citep{Wong:2011}.
To answer these questions and constrain cosmological parameters using future surveys,
theoretical predictions from numerical simulations must be accurate on smaller scales than are accessible to linear theory.
As a primary summary statistic, the matter power spectrum needs to be at percent-level precision for $k \lesssim 10 \hMpc$  \citep{Schneider:2016}.

% [History]
% first N-body and Coma cluster simulation from Peeble
% and recent developments on PM, tree, PP-PM
Modelling non-linear gravitational clustering is done using $N$-body simulations, where a dark matter fluid is sampled by macro-particles and evolved using a smoothed gravitational force.
Each macro-particle is representative of an ensemble of microscopic dark matter particles.
Generations of computational physicists have improved the accuracy of the gravitational evolution, and created quicker and more scalable algorithms to drive the mass resolution of the simulations ever higher \citep{Hockney:1988, BarnesHut:1986, Couchman:1995, Greengard:1987, Dehnen:2002}.

The mass resolution necessary to robustly predict the power spectrum at $k \sim 10 \Mpch$ pushes the computational limits of contemporary supercomputers.
To adequately sample a high-dimensional input parameter space with Markov chain Monte Carlo (\textsc{mcmc}),
millions of samples are needed, while a limited number (at best a few hundred to a few thousand) of high-fidelity simulations are computationally possible.
% [Accelerate the process of N-body]
% PKDGRAV3: trillion particles, but we not just need one but many
% simulations to explore the cosmologies and galaxy formation theories
% while these developments effectively accelerate
% the time-to-solution theoretical predictions
% They usually challenge the state-of-art supercomputer
% with the most expensive simulation setup, finest
% resolution, largest box. And that often results in
% not enough samples of simulations.
% emulation: PkANN, Mira-Titan, Cosmic Calibration, Euclid

An efficient way to perform accurate cosmological inference with a limited number of simulations is to use \textit{emulators}. Emulators are flexible statistical models, usually built with Gaussian processes, which learn the mapping from input cosmological parameters to summary statistics.
This reduces the number of costly forward simulations by effectively interpolating the function outputs.

% This allows us to effectively interpolate functional outputs and reduce the number of costly forward simulations. %They also work as cheap surrogates, allowing for fast function evaluations suitable for solving inverse problems using \textsc{mcmc}.
%%%% Moved according to referee %%%%
Emulators have been applied extensively in the field of cosmological inference.
\cite{Heitmann:2006, Habib:2007} proposed a cosmic calibration project to make percent-level predictions on the matter power spectrum using a Bayesian emulator.
\cite{Heitmann:2009, Lawrence:2010, Heitmann:2014} implemented this cosmic emulator in their Coyote Universe suite using $37$ high-resolution simulations.
\cite{Heitmann:2016, Lawrence:2017} designed the Mira-Titan Universe suite to train emulators to make precise theoretical predictions using $36$ simulations.
The latest Euclid preparation \citep{Euclid:2020} runs $250$ simulations ($3000^3$ particles) to prepare their emulator for the matter power spectrum.
Besides Gaussian processes, \cite{Agarwal:2014} used a neural network to build a cosmic emulator from $6\,380$ $N$-body simulations spanning $580$ cosmologies.

Beyond the matter power spectrum,
emulators have been trained to predict the halo mass function \citep{Bocquet:2020}, the concentration-mass relation for dark-matter haloes \citep{Kwan:2013}, the galaxy power spectrum \citep{Kwan:2015}, the galaxy correlation function \citep{Zhai:2019}, the halo bias \citep{McClintock:2019}, weak lensing peak counts \citep{Liu:2015}, the cosmic shear covariance \citep{Harnois:2019}, weak lensing voids \citep{Davies:2020}, the 21 cm signal \citep{Kern:2017}, and the Lyman-$\alpha$ 1D flux power spectrum \citep{Bird:2019}.
They also have been used for inferring beyond-$\Lambda$CDM cosmologies \citep{Giblin:2019,Pedersen:2020} and $f(R)$ gravity cosmologies \citep{Ramachandra:2020}.

While all these emulators successfully predict summary statistics using high-fidelity simulations,
one question which remains is how to minimize the number of necessary training simulations to achieve a given accuracy.
% Gaussian processes (\gp) \citep{Rasmussen05}, in combination with Bayesian modelling, enable the multi-fidelity emulation to be built within the probabilistic framework, which gives predictions with well-defined uncertainty quantification.
% The underlying assumption is the low-fidelity models can capture the general trends of the true function,
% and there exists a strong correlation between models from different fidelities, but the strengths of the correlations might be different across input space.
%%%% Moved according to referee %%%%
% multi-fidelity emulations
Here we demonstrate that building cosmological emulators from simulations can be improved with multi-fidelity models. Multi-fidelity models \citep{Kennedy:2000} minimize the computational cost by combining the predictive power of simulations at different resolutions.
%Instead of improving the sampling strategy,
They fuse the expensive but accurate \textit{high fidelity} data with cheaply-obtained \textit{low fidelity} approximations.
One standard model used by the multi-fidelity emulation is a \textit{multi-output Gaussian process} \citep{Bonilla:2007}.
A multi-output Gaussian process (multi-output {\gp}) generalizes a single-output {\gp} to multiple outputs, while building a cross-covariance function to model the shared information between outputs.
In this paper, low and high fidelity correspond to simulations at different resolutions. High-fidelity simulations have a finer mass resolution while low-fidelity simulations have a coarser mass resolution.

To train the multi-fidelity emulator using as few high-resolution simulations as possible,
we also propose a method for selecting high-fidelity training samples, based on minimizing the loss computed among the low-fidelity simulations.
% Selecting high-fidelity samples for training is particularly important when we have fewer simulations ($\leq 5$) than the dimensions of the parameters in interest ($5 - 8$).
By optimizing the low-fidelity emulator's loss, we show that one can efficiently train a multi-fidelity emulator by avoiding worst-case combinations of the high-fidelity training samples.

%Another example of low fidelity approximations is the simulation lacking specific physical models, such as baryon physics.\spb{Let's leave this out for now since we don't do it}
% cosmological usage of multi-fidelity
% limitations, assumptions (low-fidelity
% capture the general trends).
% recent developments on super-resolution and CarPool
% Yi Lin paper and CarPool paper (and the papers Yi cited)
Computational astrophysicists have used methods similar to multi-fidelity modelling to minimize the cost of performing high-resolution simulations \citep{Lukic:2015, Chartier:2020}. A notable example is Richardson extrapolation \citep{Richardson:1911}, a numerical method to improve a simulation's accuracy by combining a sequence of simulations with varied spatial resolutions and fixed cosmologies.
More recently, generative adversarial networks (GAN) have been used to produce high-resolution density fields \citep{Doogesh:2020} and particle displacements \citep{Li:2020} from low-resolution (but larger volume) input data.
%  used a super-resolution GAN to map the particle displacements from low-resolution $N$-body simulations to their high-resolution counterparts. This allows large volume simulations at low resolution to be combined with small volume but higher resolution simulations to create a single (artificial) large volume simulation at high resolution.
In principle, such `super-resolution' simulations could be implemented as a multi-fidelity emulator's high-fidelity training set, allowing an emulator to be built to a scale not directly accessible to simulations.

% Bayesian optimisation in cosmology
\cite{Rogers:2019,Leclercq:2018} proposed using Bayesian optimization to improve emulator accuracy by a sequential choice of new simulation points designed to globally optimize the emulator function.
Similar approaches to iterative selection of training data in a cosmological parameter space have been presented by \cite{Takhtaganov:2019,Pellejero:2020}.
Computer scientists and engineers, including \cite{Huang:2006,Forrester:2007,Lam:2015,MISO:2016,McLeod:2017}, have extensively studied combining multi-fidelity methods with Bayesian optimization.\footnote{\cite{BOTutorial:2018} has a subsection that provides a short review on multi-fidelity Bayesian optimization.}
Multi-fidelity Bayesian optimization arises when a cheaper approximation to the object function exists.

We present a multi-fidelity emulator for the matter power spectrum, as output by the cosmological simulation code {\mpgadget} \citep{Springel:2003, MPGADGET:2018}.
In this paper, we target percent level \textit{relative accuracy}: how well our emulators can reproduce the matter power spectra at our highest fidelity. We defer producing an emulator which allows percent level accurate reconstruction of observations or a hypothetical ideal simulation to future work. The main goal of this paper is to demonstrate that our multi-fidelity techniques can be used to reduce the computational budget required for an emulator.

% In future work we will run large simulations and build a multi-fidelity emulator converged in a specific scale and redshift range, ready for applying to observations.
% Performing $37$ high-resolution simulations requires large supercomputer resources.
We use two fidelities in a $256 \Mpch$ box:
a fast but low resolution version with $128^3$ dark-matter particles
and a slow but high resolution version with $512^3$ particles.
Even with only $3$ high-fidelity simulations and $50$ low-fidelity simulations, we show that we can predict the high-resolution matter power spectrum at percent-level accuracy on average at $k \leq 6.4 \hMpc$ at $z = 0$, with a total computational cost $\lesssim 4$ high-fidelity simulations.
Although we only show our application to the matter power spectrum, the methods presented in this paper could apply to other summary statistics, e.g.,
the halo mass function or the {\Lya} 1D flux power spectrum.

\cite{vanDaalen:2011} showed that the lack of AGN feedback affects a dark matter-only simulation significantly (compared to the error requirements of upcoming surveys) at $k > 0.1 \hMpc$. Furthermore, baryon cooling can alter the power spectrum at $k \sim 10 \hMpc$ \citep{White:2004}.
However, as techniques exist to model this effect in post-processing \citep{Schneider:2020}, we defer extending our technique to hydrodynamical simulations including AGN feedback to future work. Here we validate that a multi-fidelity emulator is useful in the simplest case: dark matter-only $N$-body simulations.

% MF emulator: standard hierarchical one
We build two types of multi-fidelity emulators. One uses the linear autoregressive model of \cite{Kennedy:2000} (first-order autoregressive model, AR1), which we will call the ``linear multi-fidelity model.''
The second multi-fidelity emulator uses the non-linear fusion model of \cite{Perdikaris:2017} (nonlinear auto-regressive Gaussian process, NARGP), and which we call the ``non-linear multi-fidelity emulator.''\footnote{AR1 and NARGP are acronyms used in \cite{Perdikaris:2017,Cutajar:2019}. In this paper, AR1 and linear multi-fidelity emulator are interchangeable, and NARGP and non-linear mutli-fidelity emulator are interchangeable.}
\cite{Kennedy:2000} model the scaling factor between fidelities as a scalar, while \cite{Perdikaris:2017} allow the scaling factor to depend on input parameters.
Our implementation of AR1 and NARGP is based on \texttt{emukit} \citep{Emukit:2019},\footnote{\url{https://github.com/EmuKit/emukit}} an open-source package for emulation and decision making under uncertainty, with the modifications mentioned above.\footnote{For a detailed comparison between AR1 and NARGP, see \cite{Cutajar:2019}. An example code for the comparison between AR1 and NARGP can be found in Emukit's examples.}

% The final product of this work will be presented as a light-weighted multi-fidelity framework for the matter power spectrum.
% Users can run very low-resolution simulations ($128^3$ particles) on their laptops to explore the parameter space and predict the high-resolution outputs before they decide to put their computational resources for a big run.

In Section~\ref{sec:simulations}, we briefly describe the simulation code, {\mpgadget}, for training the emulator.
We recap the general formalism of a single-fidelity Gaussian process emulator in Section~\ref{sec:single}.
Section~\ref{sec:multi} describes the formalism of a multi-fidelity emulator (\mfemu).
We explain our sampling strategy in Section~\ref{sec:sampling_strategy}.
Section~\ref{sec:results} shows the results, with comparisons between multi-fidelity emulation and single-fidelity emulation.
We summarize the runtime for the {\mpgadget} simulations in Section~\ref{sec:runtime}.
We conclude with a summary of key contributions and potential applications of our work in Section~\ref{sec:conclusions}.
Our code for multi-fidelity emulation in the matter power spectrum is publicly available at \url{https://github.com/jibanCat/matter_multi_fidelity_emu}.

% \section{Methods}
% \label{sec:methods}

% In this section, we first briefly describes the simulation code we performed for generating the training sets with different fidelities in Section~\ref{subsec:simulations}.
% We next describe how we train a single-fidelity emulator in Section~\ref{subsec:single}.
% Finally, in Section~\ref{subsec:multi},
% we then describe the multi-fidelity emulation framework we used.

\section{Simulations}
\label{sec:simulations}

We prepare our training set by running dark matter-only simulations using the massively parallel $N$-body code {\mpgadget} \citep{Feng:2018}.\footnote{\url{https://github.com/MP-Gadget/MP-Gadget/}}
{\mpgadget} is a publicly available $N$-body+Hydro cosmological simulation code derived from {\gadget} \citep{Springel:2003}.
It is parallelized using a hybrid OpenMP/{\mpi} strategy and
has successfully performed a hydrodynamical simulation using all $8\,032$ \textit{Frontera} nodes, a total of $449\,792$ cores,
demonstrating its good scalability properties.
The gravitational forces are computed using a Fourier transform based particle-mesh algorithm on large scales and a Barnes-Hut tree on small scales.

% {\mpgadget} uses a TreePM gravity solver, as in {\gadget}. But, for long range forces, we changed the particle-mesh gravity solver from FFTW2 to a 2-d tile FFT library, which significantly improves the performance.
% We used Fourier-space finite differencing of gravity forces, as used in HACC \citep{Habib:2014}, to reduce memory overhead.
% Further details in Bird et al (in prep).

We initialise our simulations from the linear power spectrum produced by {\class} \citep{Lesgourgues:2011} at $z=99$ using the Zel'dovich approximation
\citep{Zeldovish:1970}. The dark matter particles then evolve through gravitational dynamics. The matter power spectra are computed from the output snapshots of {\mpgadget}, and used as our emulation targets.
% more simulation descriptions moved from referee report
In this paper, we fix the IC noise in the nodes and change only the cosmology for the emulator training. We do not use the ``paired and fixed'' technique \citep{Angulo:2016}, but it would be easy to do so using only low resolution simulations as these pairings are designed to remove variance on large scales.

The matter power spectrum, $P(k)$, is a compressed summary statistic of the over-density field, $\delta(x)$, evaluated as an angle average of the Fourier-transformed overdensity field:
\begin{align}
   P(|\kvec|) &= \langle \hat{\delta}^*(k) \hat{\delta}(k)\rangle, \\
   \hat{\delta} (\kvec) &= \int \dd^3 \rvec \delta(\rvec) e^{- 2 \pi i \kvec \cdot \rvec}.
\end{align}
We measure the power spectrum with a cloud-in-cell mass assignment, which is deconvolved.
The Fourier transform is taken on a mesh the same as the PM grid of the simulation, which has a resolution of 2 times the mean inter-particle spacing.

For a multi-fidelity problem, our data are from simulations at different resolutions.
Since low resolution simulations are cheaper to obtain (but are only approximations to the high resolution results), we typically have a limited number of high-fidelity data and many low-fidelity approximations.

% table of notations
\begin{table}
   \caption{Notations and definitions}
   \begin{tabular}{|l|l|}
      Notation & Description \\
      \hline
      {\highres} & High-resolution simulation, $512^3$ particles \\
      {\lowres} & Low-resolution simulation, $128^3$ particles \\
      % simulation math notation
      $x_{i,t}$ & Input cosmological parameters at $i$th simulation\\
      & at fidelity $t$\\
      $y_{i,t}$ & Matter power spectrum at $i$th simulation\\
      & at fidelity $t$, at log scale\\
      $n_t$ & Number of simulations at fidelity $t$ \\
      % simulation convenient notation
      $\npart$ & Number of particles per box side\\
   \end{tabular}
   \label{tab:notation}
\end{table}

To make the text of this section consistent with the following sections,
we provide some notation to bridge the terminology, summarized in Table~\ref{tab:notation}.
We have data from $s$ different fidelities (simulation resolutions). For each fidelity, we have pairs of inputs and
outputs $\Data_{t} = \{x_{i,t}, y_{i,t} \} = \{ \xvec_{t}, \yvec_{t} \}$,
where $t = 1, \dots, s$ denotes the fidelity level from low to high,
and $i = 1, \dots, n_t$ where $n_t$ is the number of data pairs at fidelity $t$
and $i$ indexes each individual simulation.
The data pairs $\Data_t = \{ \xvec_t, \yvec_t \}$ for our emulation setup are the cosmological parameters
of the simulations and the power spectrum outputs.
Here we have $s=2$ for two mass resolutions:
$128^3$ and $512^3$ dark matter-only simulations.
We will denote $128^3$ as low-resolution (\lowres, $t=1$)
and $512^3$ as high-resolution (\highres, $t=2$).

Each fidelity will have a different number of simulations, $n_t$.
Practically, the number of {\lowres} simulations will be much larger than the number of {\highres} simulations, $n_1 > n_2$.
The compute time for {\lowres} ($\npart = 128$) is $\sim 20$ core hours and $\sim 2\,000$ core hours for {\highres} ($\npart = 512$).
We will empirically show we only need $3$ {\highres} and $50$ {\lowres} to train a multi-fidelity emulator with an average emulator error per $k$ smaller than $1\%$.

% since $126^3$ simulations are much more computationally cheap to obtain, even could be run on a laptop.

% Our current demonstration is conditioned on a given redshift bin $z_0$.
We do not emulate the matter power spectrum across redshifts, conditioning on a given redshift bin $z_0$. We generally focus on $z_0 = 0$, but will discuss multi-fidelity emulators at $z_0 = 1$ and $z_0 = 2$ in Section~\ref{subsec:higher_z}.

\subsection{Latin hypercube sampling}

% Since we use a squared exponential kernel to train our {\gp},
% uniformly spreading out sampling points will be preferred to minimize the worst case interpolation error.
As \cite{Heitmann:2009} mentioned,
a space-filling Latin hypercube design is well suited for {\gp} emulators of the matter power spectrum.
For a training set with $d$-dimensional inputs and $N$ simulations,
an $N^d$ grid is created first, and simulations are placed on this grid so that only one simulation is present in any row or column.
The Latin hypercube design improves on random uniform sampling by ensuring that the chosen points do not crowd together in any subspace.

We apply a Latin hypercube design on the input parameter space, $\inputparam$.
We vary the $\Lambda$CDM cosmological parameters $\{ h, \Omega_0, \Omega_b, A_s, n_s \}$, which are the Hubble parameter $h = H_0 / (100 \kms \Mpc^{-1})$, the total matter density $\Omega_0$, the baryon density $\Omega_b$, primordial amplitude of scalar fluctuations $A_s$, and the scalar spectral index $n_s$.
We use the same set of $\Lambda$CDM cosmological parameters as \cite{Euclid:2020}, allowing us to compute the relative errors of our simulations with respect to EuclidEmulator2.

We use bounded uniform priors for the input parameters:
\begin{equation}
   \begin{split}
      h        &\sim \uniform[0.65, 0.75];\\
      \Omega_0 &\sim \uniform[0.268, 0.308];\\
      A_s      &\sim \uniform[1.5\times10^{-9}, 2.8\times10^{-9}];\\
      n_s      &\sim \uniform[0.9, 0.99];\\
      \Omega_b &\sim \uniform[0.0452, 0.0492].
   \end{split}
   \label{eq:prior_volume}
\end{equation}
The dark energy density is $\Omega_{\Lambda} = 1 - \Omega_0$.
The prior volume surrounds the WMAP 9-year cosmology \citep{WMAP:2013}.
The code to handle the simulation input files and Latin hypercube design is publicly available at \url{https://github.com/jibanCat/SimulationRunnerDM}.

\subsection{Preprocessing of the simulated power spectrum}
\label{subsec:preprocessing_powerspecs}

\begin{figure}
   \includegraphics[width=\columnwidth]{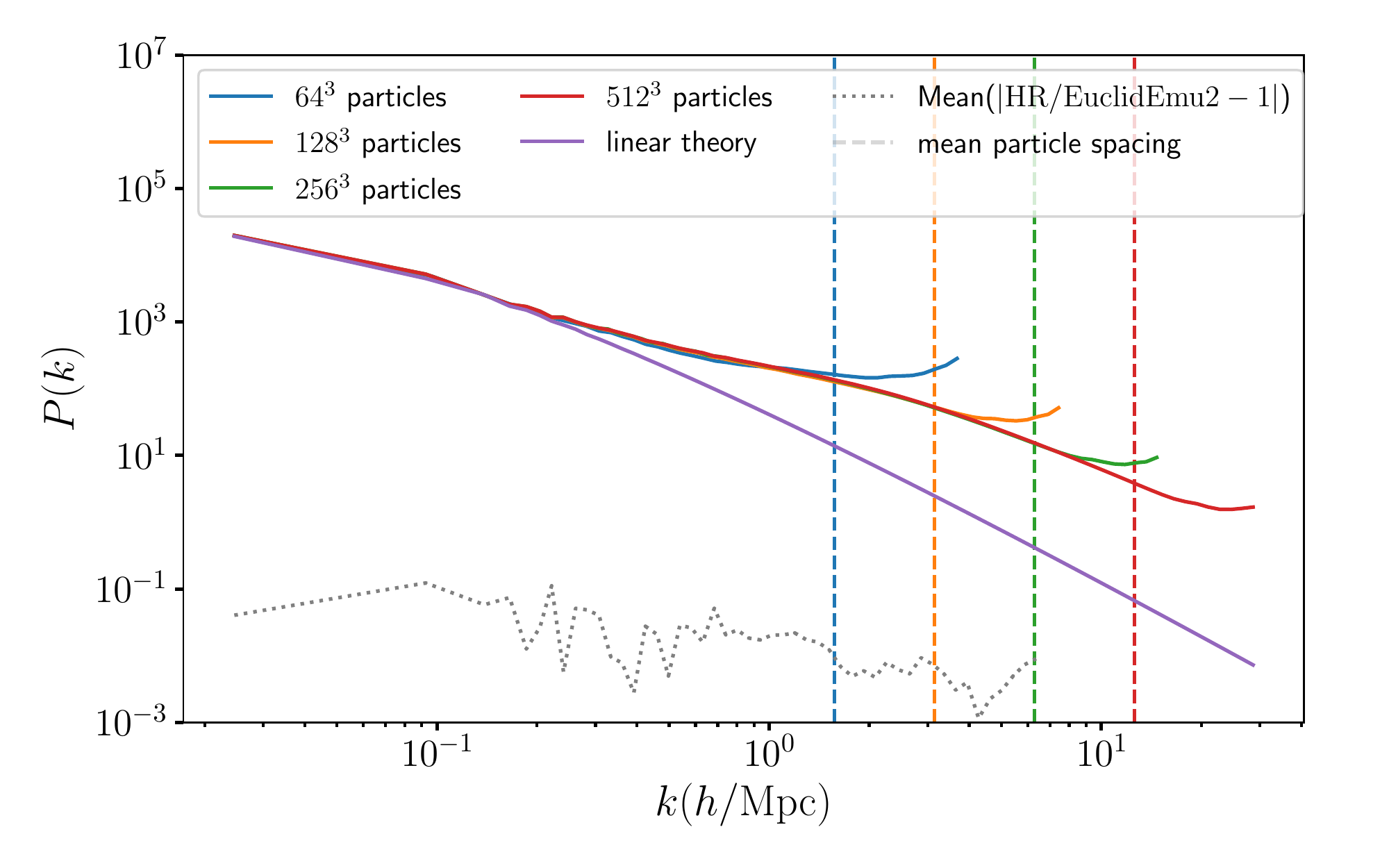}
   \caption{The matter power spectrum output by {\mpgadget} at different mass
   resolutions.
   The vertical dash lines indicate the mean particle spacing $k_\mathrm{spacing}$
   for a given mass resolution.
   \textbf{(Blue):} The matter power spectrum from a dark-matter only {\mpgadget}
   simulation with $64^3$ particles.
   \textbf{(Orange):} The matter power spectrum from {\mpgadget} with $128^3$
   particles.
   \textbf{(Green):} The matter power spectrum from {\mpgadget} with $256^3$
   particles.
   \textbf{(Red):} The matter power spectrum from {\mpgadget} with $512^3$
   particles.
   \textbf{(Purple):} Linear theory power spectrum.
   The cosmology parameters are $h = 0.675, \Omega_0 = 0.278, \Omega_b = 0.0474, A_s = 1.695\times 10^{-1}, n_s = 9.405 \times 10^{-1}$.
   The dotted line shows the relative error of {\highres} ($512^3$ simulations) compared with EuclidEmulator2 \citep{Euclid:2020},
    averaged over four different cosmologies.
   }
   \label{fig:particle_spacing}
\end{figure}

% Thus, the robustness of the small-scale physics relies on how small the grid size of a simulation can achieve.
% Modelling the unphysical parts of the small-scale power spectrum will cause the difficulties during training the {\mfgp} and also potentially lead to dubious predictions.
A numerical simulation is constrained by its box size and number of particles.
The mass resolution limits the smallest scale (the highest $k$) of the power spectrum.
Thus, high-fidelity simulations can model smaller scales, not fully resolved in low-fidelity simulations, as shown in Figure~\ref{fig:particle_spacing}.
% the low-fidelity power spectrum is shorter than the high-fidelity counterpart.

For $k$ larger than the mean particle spacing, $P(k)$ differs substantially from the resolved value, due to artifacts of the macro-particle sampling.
The scale of the mean particle spacing is
\begin{equation}
   k_\mathrm{spacing} = 2 \pi \frac{\npart}{\boxsize},
   \label{eq:ptl_spacing}
\end{equation}
where $\npart$ is the number of particles per side of the box.
For instance, if we have $512^3$ particles in the box, then $\npart = 512$.
$\boxsize$ is the size of the simulation box in units of $\Mpch$.

% [todo] an example plot to show we cut off the particle spacing scale.
% at the same time showing the data pairs from different fidelities
%  | (left) parameter selection | (right) power spec with diff fidelities |

% describe that we keep the low-fidelity data below the initial particle spacing.
We use the same set of matter power spectrum $k$ bins for all fidelities.
The available information at small scales is sparse for the low-fidelity spectrum.
To resolve the issue, we fix the $k$ bins to high fidelity
and linearly interpolate the low-fidelity power spectrum in a $\log_{10}$ scale, $\log_{10}{P(k)}$, onto the high-fidelity $k$ bins.
The maximum $k$ is set to be $\simeq 6.4 \hMpc$ when using $\npart = 128$ as our low-fidelity training set.
However, in practice we found that $128^3$ and $512^3$ simulations shared similar $k$ bins with small offsets at small scales.

We do not model the high-fidelity spectrum with $k$ larger than the maximum $k$ of the low-fidelity spectrum:
\begin{equation}
   \max{k_{t = 2}} = \max{k_{t = 1}},
\end{equation}
where $t$ indicates the fidelity level and $t = 2$ is the highest fidelity.
If we do not have any data at a given $k$ from low-fidelity,
we cannot extract the correlations between fidelities without other more
significant assumptions.
In other words,
the maximum $k$ we can model is limited by the data available from the low-fidelity simulations,
which always have a lower maximum $k$ than high-fidelity simulations.
We note that it is possible to get a higher maximum $k$ by particle folding or by increasing the size of the PM grid size used for estimating the power spectrum, although we do not do that here.

We do model the low-fidelity $P(k)$ even on scales smaller than the mean particle spacing, $k > k_\mathrm{spacing}$.
We made this particular decision because we have a prior belief that even though $P(k > k_\mathrm{spacing})$ is highly biased,
it still captures some information about how $P(k)$ depends on cosmological parameters.
Thus, we should be able to exploit the correlations between fidelities and improve the emulator accuracy at those scales.

To summarize, we:
\begin{enumerate}
   \item Use the same set of $k$ bins across different fidelities.
   \item Preserve all available $P(k)$ from low-fidelity, even scales smaller than the simulation's mean particle spacing.
\end{enumerate}

% \begin{figure}
%    \includegraphics[width=\columnwidth]{images/plot_mf_outputs_single.pdf}
%    \caption{An example of a pair of low-fidelity and high-fidelity power spectrum for training our {\mfemu}.
%    \textbf{(Blue):} The matter power spectrum with $128^3$ particles.
%    \textbf{(Orange):} The matter power spectrum with $512^3$ particles, with the maximum $k$ to be the same as the low-fidelity counterpart.
%    }
%    \label{fig:chopped_outputs}
% \end{figure}

% \spb{Do we need Fig.\ref{fig:chopped_outputs}?}
% We show an example plot of our low-fidelity and high-fidelity power spectrum in Figure~\ref{fig:chopped_outputs}.

\section{Single-fidelity emulators}
\label{sec:single}

Here we briefly recap how we train a single-fidelity emulator. Readers familiar with this material may wish to skip to Section~\ref{sec:multi}.
The notation we use in this section follows those of \cite{Perdikaris:2017,Cutajar:2019}.
Consider a supervised learning problem,
in which we wish to learn the mapping relation, $f$, between a set of input and output pairs
$\Data = \{x_i, y_i\} = \{\xvec, \yvec\}$,
where $i = 1, \dots, n$:
\begin{equation}
   y = f(x), ~\textrm{ with}~ x \in \realspace^d,
\end{equation}
where $d$ is the dimension of the input space.
A Gaussian process (\gp) \citep{Rasmussen05} is a probabilistic framework modelling the observations, $\yvec$, as drawn from a noisy realization of a single random function $f$ with a likelihood $p(\yvec \mid f)$.
It models the distribution over $f$
\begin{equation}
   p(f) = \GP(f; \mu, K),
\end{equation}
with $\mu$ the {\gp} mean prior function, which is usually assumed to be a zero mean prior, and $K$ the covariance kernel function specified by a vector of hyperparameters, $\thetavec$.
For a given set of inputs, $x_1, x_2, \dots, x_n$, the kernel function evaluated on these points produces a symmetric, positive-definite covariance matrix
$K_{ij} = K(x_i, x_j; \thetavec)$ with $K \in \realspace^{n\times n}$.

The choice of the covariance kernel depends on our prior knowledge about the data.
The hyperparameters of a chosen kernel are optimized by maximizing the marginal log-likelihood:
\begin{equation}
   \log p(\yvec \mid \xvec, \thetavec)
   = - \frac{1}{2}\log|\Kvec|
   - \frac{1}{2}\yvec^{\top}\Kvec^{-1}\yvec
   - \frac{n}{2}\log2\pi.
   \label{eq:mle}
\end{equation}

For an emulator,
the main purpose is to predict an output $f_* = f(x_*)$ from a new input point $x_*$, given the provided data $\Data$.
% That means we want to make predictions on a new output $f_*$, conditioned on a new input $x_*$ and previous data $\Data = \{\xvec, \yvec\}$:
\begin{equation}
   \begin{split}
      p(f_* \mid \Data, x_*) &= \normal(f_* \mid \mu_*(x_*), \sigma^2_*(x_*) ), \\
      \mu_*(x_*)             &= \kvec_{*n} \Kvec^{-1} \yvec, \\
      \sigma^2_* (x_*)       &= K(x_*, x_*) - \kvec_{*n} \Kvec^{-1} \kvec_{*n}^\top,
   \end{split}
   \label{eq:sf_posterior_pred}
\end{equation}
where $\mu_*$ is the posterior mean and $\sigma_*$ is the standard deviation of the uncertainty in the estimate of the predictions.
The vector $\kvec_{*n}$ is the covariance between the new point and trained data,
$\kvec_{*n} = [K(x_*, x_1), \dots, K(x_*, x_n)]$.

\subsection{Cosmological emulators}
\label{subsec:gp_emulator}

% Here we briefly describe how a {\gp} connects to an emulator.
Consider we have a set of dark matter-only simulations with fixed box size and mass resolution.
At each redshift bin $z_0$,
we can compute the matter power spectrum, $P(k, z=z_0)$, given a set of input parameters.
We will use the log power spectrum, $\log_{10}{P(k, z=z_0)}$, as our training data.

% Input parameters are chosen depending on the science goals of the experiment.
% The goal of this paper is to demonstrate the multi-fidelity emulation in cosmology,
% so these five parameters were chosen solely to show the multi-fidelity emulation could extrapolate the power spectrum to small scales,
% even though the input cosmologies varied a lot.

The training data, $\Data = \{x_i, y_i\}$, are defined as
\begin{equation*}
   \begin{split}
      x_i &=  [ h_i, {\Omega_0}_i, {\Omega_b}_i, {A_s}_i, {n_s}_i];\\
      y_i &= \log_{10}{ P(k, z=z_0)},
   \end{split}
\end{equation*}
where $i = 1, \dots, n$ indicates the $i$th simulation we run with this specific set of input parameters.
% [post-referee] I feel this part is repetitive
% For a single simulation, the matter power spectrum $P(k, z=z_0)$ will be binned into several $k$ modes.
% Each input parameter set $\{ h, \Omega_0, \Omega_b, A_s, n_s \}$ will generate multiple outputs corresponding to different $k$ modes.

The rest of the modelling is choosing an appropriate covariance function $K(x, x')$.
We use a squared exponential kernel and use automatic relevance determination ({\ard}) weights for each input dimension.
{\ard} assigns each input dimension, $x_i$, a separate hyperparameter, $w_i$:
\begin{equation}
   \begin{split}
      K(x,x'; \thetavec)
      =
      \sigma^2 \exp{
      \left(
         - \frac{1}{2}
         \sum_{i=1}^{d}
         w_i (x_i - x_i')^2
      \right)}
   \end{split}
   \label{eq:single_fidelity_emu_K}
\end{equation}
where $i = 1, \dots, d$ indicates the dimension of the input space $x \in {\realspace}^d$.
$\sigma^2$ is the variance parameter for the squared exponential kernel, $\{w_i\}_{i=1}^d$ are the {\ard} weights.
$\{w_i\}_{i=1}^d$ are inverse length scales, which define the degree of smoothness at a given input dimension.
We note that we assign independent hyperparameters, $\thetavec = \{\sigma^2, w_1, \dots, w_d\}$, for each $k$ mode.\footnote{\cite{Takhtaganov:2019} refers to this approach as the many single-output approach (MS).}
A larger $w_i$ corresponds to a smaller length scale, reflecting that the learned function varies more in the $i$th dimension.
On the other hand, a smaller $w_i$ implies a larger length scale, indicating that the learned function is smoother along the $i$th dimension.
{\ard} allows each dimension of the learned function to have a different degree of smoothness.

We do not decompose the power spectrum into principle components for training the emulators, as described by \cite{Heitmann:2006,Habib:2007} because we want to compare single-fidelity emulators to the multi-fidelity emulators, and an {\mfemu} only has a limited number of high-resolution simulations available.
In our default case, we only have $3$ high-resolution simulations for an {\mfemu}, and it is not sensible to perform dimension reduction on three power spectra.

To ensure that our single-fidelity emulator is not unfairly disadvantaged in the comparison with our multi-fidelity emulator by poorly constrained hyperparameters, we built a single-fidelity emulator which shared kernel parameters across all $k$ modes and empirically verified that it had similar performance.

% the main section for the multi-fidelity emulation
% [todo] some plots need to be prepared:
% 1) the mean prior for the auto-regressive model, before and after the conditioned
% 2) covariance: not sure how to plot it but it's necessary to make the
%    the article to be explicit
% 3) Data plots: power spectra for different fidelities
\section{Multi-fidelity emulator}
\label{sec:multi}

In this section, we describe how we train a multi-fidelity emulator.
We outline the modelling assumptions in Section~\ref{subsec:assumptions}.
Section~\ref{subsec:linear_mf} describes the formalism of the linear multi-fidelity emulator proposed by \cite{Kennedy:2000}, a multi-output {\gp} with a linear correlation between fidelities.
Section~\ref{subsubsec:non_linear_mf} outlines the non-linear multi-fidelity emulator of \cite{Perdikaris:2017}, which models the correlation between fidelities as a function of cosmological parameters. We follow the notation and formalism of \cite{Kennedy:2000,Perdikaris:2017,Cutajar:2019}.

\subsection{General assumptions}
\label{subsec:assumptions}

Here we outline our modelling assumptions, following the assumptions made in \cite{Kennedy:2000}:
\begin{enumerate}
   \item \textbf{Correlations between the code fidelities:} For an $N$-body simulation,
   the simulation cost depends on the mass resolution.
   We assume a simulation with a low mass resolution can approximate a simulation with a high mass resolution.
   The matter power spectrum from different fidelities is strongly correlated at large scales since all fidelities are resolved and the mass resolution has negligible effects.
   At small scales, however, we expect different fidelities are only weakly correlated.
   \item \textbf{Smoothness:} For an emulation problem, we assume that neighbouring inputs give similar outputs.
   For example, suppose two sets of input parameters to {\mpgadget} are close to each other.
   In that case, we assume that an $N$-body simulation will provide a similar outcome.
   \item \textbf{The prior belief on each fidelity is a Gaussian process:} We assume a prior belief that the mapping from code input to output is a Gaussian process for each fidelity.
\end{enumerate}

The first assumption is the core assumption of a multi-fidelity emulator.
Different levels of the same code are simulating the same physical reality.
It is thus reasonable to assume different code fidelities should correlate at some level.
However, a naive simulation, for example, $\npart = 16$ could only barely approximate a {\highres} with $\npart = 512$.
Therefore, we should also assume the correlation between fidelities depends on the distance between two fidelities in the dimension of mass resolution.

There is thus a trade-off between the strength of correlation and the computational expense: for example, a simulation with $\npart = 256$ provides more information about a {\highres} ($\npart = 512$), but running a $256^3$ simulation is $8$ times most expensive than running a {\lowres} ($\npart = 128$).

One can select an optimal choice of simulation cost by balancing the computational time and the emulation accuracy.
Here we choose $\npart = 128$ for our low-fidelity simulations because:
\begin{enumerate}
   \item The maximum $k$ is $\simeq 6.4 \hMpc$, which includes enough non-linear scales to test the emulation accuracy;
   \item A $128^3$ simulation is $64$ times cheaper than a {\highres},
   and thus the resolution difference between $\npart = 128$ and $\npart = 512$ is large enough to demonstrate whether  simulations with lower costs can accelerate the training of an emulator.
\end{enumerate}
%We mainly use $\npart = 128$ for our low-fidelity training simulations throughout the paper.
In Section~\ref{subsec:change_lowres}, we will show our method is applicable to simulations with different resolutions, $\npart= 64$ and $256$.
Empirically, we found that using $\npart = 256$ as low-fidelity is similar to $\npart = 128$, while $\npart = 64$ gives a worse emulation accuracy.
% We chose our {\lowres} based on our empirical knowledge on running {\mpgadget},
% but we encourage users optimise their choices based on the accuracy of a multi-fidelity
% emulator with a few runs of different code complexities.

The second assumption, the smoothness assumption, is the general assumption of
a {\gp} emulator.
A {\gp} emulator will have poor accuracy if the code does not behave similarly with
similar input.
The smoothness assumption is also the assumption behind the Latin hypercube sampling
scheme \cite[for a detailed discussion, see][]{Heitmann:2009}.

% The third assumption on each code level could be modelled by a {\gp} prior is our modelling decision.
A multi-fidelity emulation could in principle be implemented using other models (see \cite{Peherstorfer:2018} for different data-fit models for surrogates).
We chose to use {\gp}s simply because their Bayesian approach supports uncertainty quantification and there is a well-developed community around {\gp} emulation.

\subsection{Linear multi-fidelity emulator (AR1)}
\label{subsec:linear_mf}

We have multi-fidelity data $\Data_t$ as described in Section~\ref{sec:simulations}.
A multi-fidelity emulator is essentially inferencing the highest fidelity model conditioned on data from all model fidelities.
The final goal of a multi-fidelity emulator is to find a mapping relation $f$ such that, from an arbitrary input vector $x_{*}$, we can always find the highest fidelity code output:
\begin{equation}
   y_{s,*} = f(x_{*}).
\end{equation}
As described by \cite{Kennedy:2000},
a linear autoregressive model can be applied in a multi-fidelity setting by assuming a hierarchical order between fidelities:
\begin{equation}
   f_t(x) = \rho_{t}\, f_{t-1}(x) + \delta_t(x),
   \label{eq:linear_mf}
\end{equation}
where $f_t$ is the function emulated by a {\gp} at $t$ fidelity and $f_{t-1}$ is the function emulated at the previous fidelity level $(t-1)$.
% $\rho_t$ is a constant scaling factor that describes the correlation between the function outputs $y_t$ and $y_{t-1}$.
The linear component of Eq~\ref{eq:linear_mf} is $\rho_t$, which models the correlation between fidelities as a linear relation.
$\delta_t$ is a {\gp} modelling the bias term:
\begin{equation}
   \delta_t \sim \GP(\mu_{\delta_t}, K_t).
\end{equation}
% [modification] subtract the mean of lowRes
We modify Eq~\ref{eq:linear_mf} so inference is performed on each $k$ bin independently. For $k = \kbin$, we have independent kernel and scaling parameters for each $k = k_j$ mode.
For simplicity, we will drop the $k = k_j$ notation in the rest of the paper:
\begin{equation}
   f_t(x) = \rho_t (f_{t-1}(x) - \mu_{t-1}) + \delta_t(x).
\end{equation}
The mean of the bias term, $\mu_{\delta_t}$, is assumed to be the zero function.
For the low-fidelity part, we subtract the sample mean of the logarithm training power spectra, $\log_{10} P(k)$, and model the low fidelity part of the power spectra as a zero mean {\gp}:
\begin{equation}
   (f_{1}(x) - \mu_{1}) \sim \GP(0, K_1(x_1, x_1'; \thetavec_1)).
\end{equation}
As shown in Figure~\ref{fig:particle_spacing},
the low-fidelity power spectrum is biased high.
We pass variations of the low-fidelity power spectrum around its mean to the next fidelity to avoid passing biased outputs.
In practice, we found this slightly improves emulation accuracy for multi-fidelity models.

For the highest fidelity bias function, $\delta_s(x)$, we model the power spectrum using  a zero mean {\gp} without subtracting the sample mean. We do not have enough points at the highest fidelity for the sample mean to be a good estimate of the true mean.
Except for $t = 1$, $f_t(x)$ is completely determined by $f_{t-1}(x)$, $\delta_t(x)$, and $\rho_t$.

As mentioned by \cite{Kennedy:2000},
there is a Markov property implied in the covariance structure of Eq~\ref{eq:linear_mf}:
\begin{equation}
   \cov{\left\{ f_t(x), f_{t-1}(x') \mid f_{t-1}(x) \right\}} = 0,
   \label{eq:markov}
\end{equation}
which is true for all $x \neq x'$.
Eq~\ref{eq:markov} indicates that if we have $f_{t-1}(x)$, then other input parameters $f_{t-1}(x')$ do not contribute to training $f_t(x)$.

The Markovian property also suggests that an efficient training set $\{\Data_{1}, \Data_{2}, \cdots, \Data_{s}\}$ for a multi-fidelity {\gp} is a nested structure:
\begin{equation}
   \xvec_1 \subseteq \xvec_2 \subseteq \cdots \subseteq \xvec_s.
\end{equation}
The above notation says that, given an input point $x$ at fidelity $t$,
there must be an input $x$ in its lower fidelity $u$,
where $u < t$ and $t, u \in \{1, 2, \cdots, s\}$.
The reason for using a nested experimental design is that
since we have $\xvec_{t - 1} \subseteq \xvec_t$,
we can immediately get an accurate posterior $f_{t-1}(x)$ at the $x$ location without interpolating at the $t - 1$ level.
However, in practice we found our multi-fidelity emulators performed well even without a nested design in the input space.\footnote{Without a nested design in input space, we found, for a multi-fidelity emulator using 50 {\lowres} and 3 {\highres}, the non-nested one is only $5\%$ worse than the nested one on the relative errors.}

% Not confident here. Emukit's implementation is Kennedy:2000's implementation
% So not necessary to mention the optimization except we want to write
% Kennedy:2000's section 2.5.
%
% \cite{Perdikaris:2017} mentioned that
% the classical \cite{Kennedy:2000} model could be learned by a recursive inference proposed by \cite{Gratiet:2014},
% where replace the GP $f_{t-1}(x)$ from the previous level using GP posterior $f_{*t-1}(x)$ under nested data structure design.

% The section requires reading the recursive co-kriging paper, which would take more time
% to read since the paper has more complicate notations.
% But the conclusion is nest structure is not necessary, but having nested structure would
% reduce the hyperparameter optimisation to a simple MLE.
% \cite{Kennedy:2000, Perdikaris:2017} employ a nested experimental design to improve numerical efficiency.
% For a computer model with different costs,
% a nested experimental design $\{\Data_1, \Data_2, \cdots, \Data_s\}$  corresponds to a structure:
% \begin{equation*}
%    \xvec_1 \subseteq \xvec_2 \subseteq \cdots \subseteq \xvec_s,
% \end{equation*}
% which means the input from a higher fidelity is a subset of the input from a lower fidelity:
% \begin{equation}
%    \xvec_{t - 1} \subseteq \xvec_t,
% \end{equation}
% where $t = 1, 2, \cdots, s$ are the fidelities, with $s$ the highest fidelity.

At a given fidelity $t$, the posterior at a test input $x_{*}$ could be written as
\begin{equation}
   \begin{split}
      p(f_{*t} \mid \Data, x_{*})
      = \normal(f_{*t}; \mu_{*t}(x_*), \sigma_{*t}^2(x_*)) \,,
   \end{split}
\end{equation}
where we denote predictions from new inputs as subscript~$*$.
The predictive mean and variance are
\begin{equation}
   \begin{split}
      \mu_{*t} =& \rho_t \cdot \mu_{*{t-1}}(x_*)
      +\mu_{\delta_{t}}
      \\&+ \kvec_{*n_t} \Kvec_{t}^{-1}
       [\yvec_{t} - \rho_t \cdot \mu_{*{t-1}}(\xvec_t)
      - \mu_{\delta_{t}}];
      \\
      \sigma_{*t}^2 =& \rho_t^2 \cdot \sigma_{*{t-1}}^2(x_*)
      + K(x_*, x_*) -
      \kvec_{*n_t} \Kvec_{t}^{-1}\kvec_{*n_t}^{\top},
   \end{split}
\end{equation}
where
$\kvec_{*n_t} = [K_t(x_*, x_1), \dots, K_t(x_*, x_{n_t})]$ is a vector of covariance between the new location and the training locations at fidelity $t$.
$K_t = K_{t}(\xvec_{t}, \xvec_{t}')$ is the covariance matrix of training locations at fidelity $t$.

\subsubsection{Covariance kernel}
\label{subsubsec:covariance_kernel}

For a linear multi-fidelity emulator, we place an independent squared exponential kernel on each $\kbin$.
The mathematical form of the kernel is the same as Eq~\ref{eq:single_fidelity_emu_K}.
% \begin{equation}
%    \begin{split}
%       K_t(x,x'; \thetavec_{t})
%       =
%       \sigma_{t}^2 \exp{
%       \left(
%          - \frac{1}{2}
%          \sum_{\ell=1}^{d}
%          w_{t,\ell} (x_\ell - x_\ell')^2
%       \right)}.
%    \end{split}
%    \label{eq:rbf}
% \end{equation}
% $\sigma_{t}^2$ is the variance, and $w_{t,\ell}$ is the {\ard} weights at $\ell$th input dimension at fidelity $t$.

Having {\ard} weights means we assign different length scales to each dimension so that the kernel can be trained anisotropically.
We found that using {\ard} in the highest fidelity did not improve the model's accuracy.
Thus, we decided to assign an isotropic kernel for $\delta_s$.
For a two-fidelity emulator ($s = 2$),
we have $6$ hyperparameters in low-fidelity for each $k$ bin; $5$ of them are the length scale parameters and $1$ is the variance parameter.
We have $3$ hyperparameters for each $k$ bin in high fidelity,
with one scale factor $\rho_t$ between fidelities, one variance parameter, and one length scale parameter.
We have $49$ bins in $k$, so the total number of trainable hyperparameters is $441$.

% [learned scale factor]
\begin{figure}
   \includegraphics[width=\columnwidth]{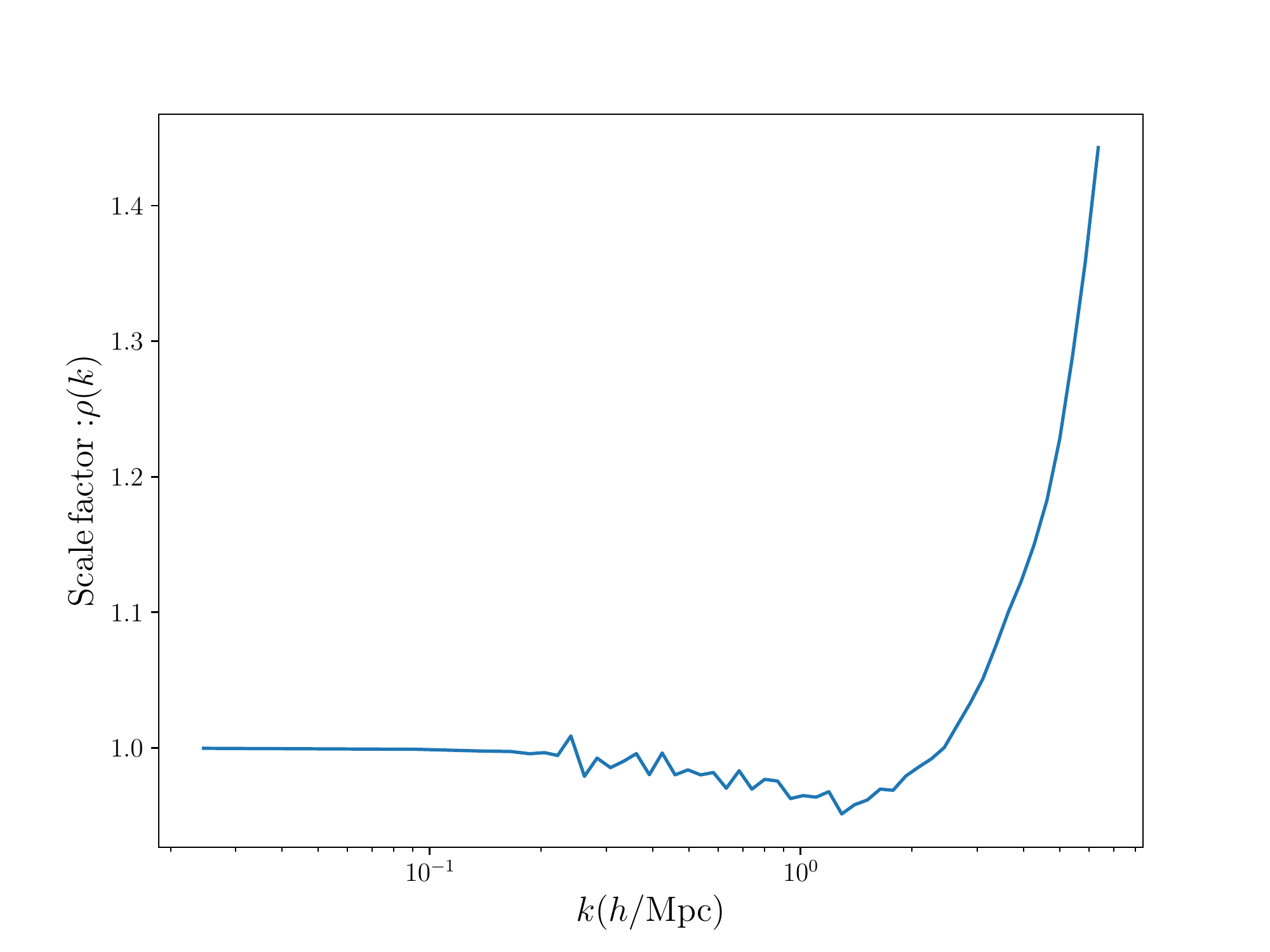}
   \caption{The learned scale factor between fidelities in the linear multi-fidelity model, $\rho$, as a function of $k$.
   This scale factor is learned from $50$ low-fidelity simulations and $3$ high-fidelity simulations.}
   \label{fig:scale_factor}
\end{figure}

Figure~\ref{fig:scale_factor} shows the learned scale factor, $\rho$.\footnote{The multi-fidelity scale factor shown Figure~\ref{fig:scale_factor} is $\rho_2$, which is $\rho_t$ when $t=2$. For simplicity, we use $\rho$ to refer to $\rho_2$ for our multi-fidelity emulators.}
$\rho$ is roughly unity at large scales $k \leq 2 \hMpc$,
but its value increases dramatically after $k > 2 \hMpc$.
Non-linear physics becomes important and the low-fidelity simulations become less reliable at small scales, making the relationship between fidelities non-trivial.
We want to emphasize that the scale factor, $\rho$, is learned from the multi-fidelity emulator. We did not enforce $\rho$ to be a specific shape during the training. Because we learn the mapping from {\lowres} to {\highres} using the training data, it is expected that {\lowres} runs deviate from {\highres} power spectra. The purpose of multi-fidelity emulation is to correct these deviations.

% \subsubsection{Learning the hyperparameters}
% \label{subsubsec:hyperparameters}

\subsection{Non-linear multi-fidelity emulator (NARGP)}
\label{subsubsec:non_linear_mf}

% As in Section~\ref{subsec:linear_mf}, we have simulated data
% $\Data_t = \{ x_{i,t}, y_{i,t} \} = \{ \xvec_{t}, \yvec_{t} \}$ for fidelity $t = 1, \dots, s$ and simulations $i = 1, \dots, n_t$.
The linear multi-fidelity model in Eq~\ref{eq:linear_mf} assumes the scale factor $\rho_t$ is independent of input parameters, $x$, and so does not model the cosmological dependence of the scale factor $\rho_t$. The non-linear multi-fidelity model proposed by \cite{Perdikaris:2017} drops this assumption, allowing the scale factor, $\rho_t(\cdot)$, to be a function of both input cosmology and output from the previous fidelity.
As for the linear multi-fidelity model, we model the non-linear multi-fidelity {\gp} independently for each $k$:
\begin{equation}
   \begin{split}
      f_t(x) = \rho_t(x, f_{t-1}(x) - \mu_{t-1}) + \delta_t(x),
   \end{split}
   \label{eq:nonlinear_mf}
\end{equation}
where $\rho_t(\cdot)$ is a function of both input parameters $x$ and the previous fidelity's output.
$\rho_t(\cdot)$ is modelled as a {\gp}.
Eq~\ref{eq:nonlinear_mf} results in a more complicated distribution over $f_t$,
a deep Gaussian process \citep{Damianou:2013}.
To avoid added computational and statistical complexity,
we follow the same approximation as \cite{Perdikaris:2017} and replace $f_{t-1}$ in Eq~\ref{eq:nonlinear_mf} with its posterior, $f_{* t - 1}$.
The result is a regular Gaussian process,
\begin{equation}
   \begin{split}
      f_t \sim \GP(0, K_{t}),
   \end{split}
   \label{eq:gt_kernel}
\end{equation}
whose kernel can be furthermore decomposed:
\begin{equation}
   \begin{split}
      K_{t}(x, x') = &K_{t_\rho}(x, x'; \thetavec_{t_\rho})
      \cdot K_{t_f}(  f'_{*t-1}(x), f'_{*t-1}(x'); \thetavec_{t_f})\\
      &+     K_{t_\delta}(x, x'; \thetavec_{t_\delta}),
   \end{split}
   \label{eq:nonlinear_factorisation}
\end{equation}
where $f'_{*t-1} \equiv f_{* t-1}(x) - \mu_{t-1}$ for simplicity.
The first kernel $K_{t_\rho}$ models the cosmological dependence of the scale factor $\rho$.
Next, $K_{t_f}$ models the covariance of the output passing from the previous fidelity to the current level.
The final term $K_{t_\delta}$ models the model discrepancy between fidelities.
For the lowest fidelity, the matter power spectrum is only modelled with $K_{t_\delta}$.

Each kernel in Eq~\ref{eq:nonlinear_factorisation}, $(K_{t_\rho}, K_{t_f}, K_{t_\delta})$, is modelled as a squared exponential kernel.
Suppose we assign a different length scale parameter for each $x$ dimension.
$K_{t_\rho}$ will have $d + 1$ hyperparameters, $K_{t_f}$ will have $2$ hyperparameters,
and $K_{t_\delta}$ will have $d + 1$ hyperparameters.
As for the linear emulator, we found no improvement in accuracy in practice by using {\ard} for the high-fidelity model.
Thus, we have $2$ hyperparameters for each kernel in high fidelity and $d + 1$ hyperparameters for low-fidelity.
To be explicit, in the high-fidelity model,
$K_{t_\rho}$ has $2$ hyperparameters, $K_{t_f}$ has $2$ hyperparameters,
and $K_{t_\delta}$ has $2$ hyperparameters.
For $d = 5$, we have $6$ hyperparameters for low-fidelity and $6$ for high-fidelity models at each $k$ bin.

% Due to the non-linear design in Eq~\ref{eq:gt_kernel},
% the posterior predictive $f_{*t-1}(x_*)$ will be non-Gaussian after $t > 2$.
% This means that if we want to design a 3-fidelity model, we have to propagate
% the uncertainty with an approximated method such as Monte Carlo integration:
% \begin{equation}
%    \begin{split}
%       p(f_{*t}(x_*))&\\
%       = \int &p( f_t(x_*, f_{* t-1}(x_*) ) \mid \Data_t, x_* )
%       p(f_{* t-1}(x_*)) \dd x_*,
%    \end{split}
% \end{equation}
% where we need to approximate the posterior predictive at $t$ level through
% uncertain inputs $x_*$.
% % here need to check the code if it is doing what they are saying

\subsubsection{Halo Model Interpretation}
\label{subsubsec:halo_model}

% cosmological interpretation using halo model
The formulation of our multi-fidelity emulator bears a marked resemblance to the equations which form the basis of {\small HALOFIT} \citep{Smith:2002dz}, and are themselves motivated by the halo model \citep{Peacock:2000qk, Seljak:2000gq}. This correspondence allows us to provide a physical interpretation of our results.
In the halo model, matter clustering is schematically divided into two components: a two-halo term and a one-halo term. The two-halo term arises from correlations between halos on largre scales, while the one-halo term, which has a weaker dependence on cosmology, is sensitive to the density profile inside each halo. We can model this by splitting the non-linear power spectrum
\begin{equation}
 P_{NL}(k)  = P_\mathrm{Q}(k) + P_\mathrm{H}(k)\,.
 \label{eq:halofit}
\end{equation}
The quasilinear term $P_\mathrm{Q}(k)$ is a two-halo term, while $P_\mathrm{H}(k)$ is a one-halo term. The two-halo term can be modelled by the linear theory power spectrum filtered by a window function $W(M, k)$:
\begin{equation}
 P_{Q}(k)  = P_\mathrm{L}(k) \left(\int W(M, k) \dd M\right)^2\,.
 \label{eq:quasilinear}
\end{equation}
The window function depends on the halo mass function and halo bias, encodes how virialisation displaces the linear matter field, and tends to unity on large scales.

There is a clear connection between this model and the form of our multi-fidelity emulator. Eq~\ref{eq:linear_mf} (AR1) and Eq~\ref{eq:nonlinear_mf} (NARGP) move between fidelities via two terms: a scaling factor $\rho$ and an additive factor $\delta_t$.
The correlations between fidelities are strong on large scales, and so $\rho \to 1$ as $k \to 0$. $\rho$ is analogous to the quasilinear window function, except that it filters not the linear theory power spectrum $P_\mathrm{L}$, but the low-fidelity $N$-body model $f_{t-1}(x)$. In the context of the halo model, it extrapolates the existing quasilinear halo filtering to include lower mass halos not included in the low-fidelity simulation.
%As the Gaussian process based machine learning we employ here does not include an a priori prescription for halo filtering, the low fidelity model must include some non-linearities.

The additive factor $\delta_t$, which is important on small scales, is analogous to the one-halo term. It models the difference in halo shot noise and internal halo profiles between resolutions. Notice that $\delta_t$, like the one-halo term, depends only weakly on cosmology, as evidenced by it requiring only one length-scale hyperparameter.

%Indeed, the reason our technique is able to increase accuracy using only a small number of high fidelity simulations is that both $\rho$ and $\delta_t$ are weakly cosmology dependent.

%For our multi-fidelity emulators in Eq~\ref{eq:linear_mf} (AR1) and Eq~\ref{eq:nonlinear_mf} (NARGP), the correlations between fidelities are strong on large scales, and the two-halo term will be learned by the scaling factor $\rho$.
%On small scales,
%the bias $\delta(\cdot)$ is used to compensate for the discrepancy between fidelities.
%The bias function is essentially calibrating the one-halo term using high-fidelity simulations.

\section{Sampling Strategy for High-Fidelity Simulations}
\label{sec:sampling_strategy}

In this section, we will describe how we select the training simulations for our multi-fidelity emulators.
We will first describe the nested structure implemented in multi-fidelity emulators in Section~\ref{subsec:nested_training_sets}.
Section~\ref{subsec:optimze_loss_low_fidelity} explains how we find the optimal choice of high-fidelity training simulations.

\subsection{Nested training sets}
\label{subsec:nested_training_sets}

% [nested structure input plot]
\begin{figure*}
   \includegraphics[width=\columnwidth]{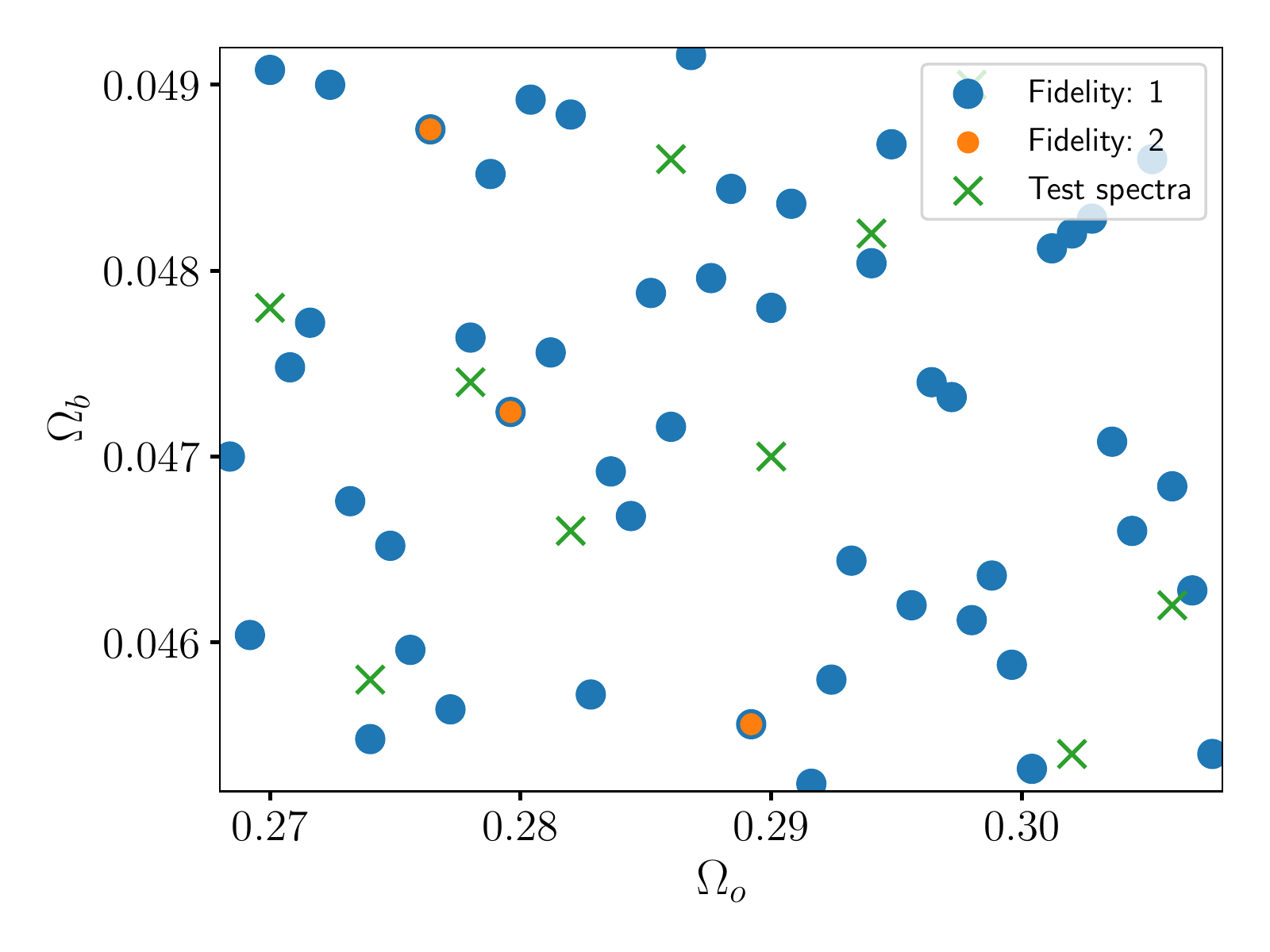}
   \includegraphics[width=\columnwidth]{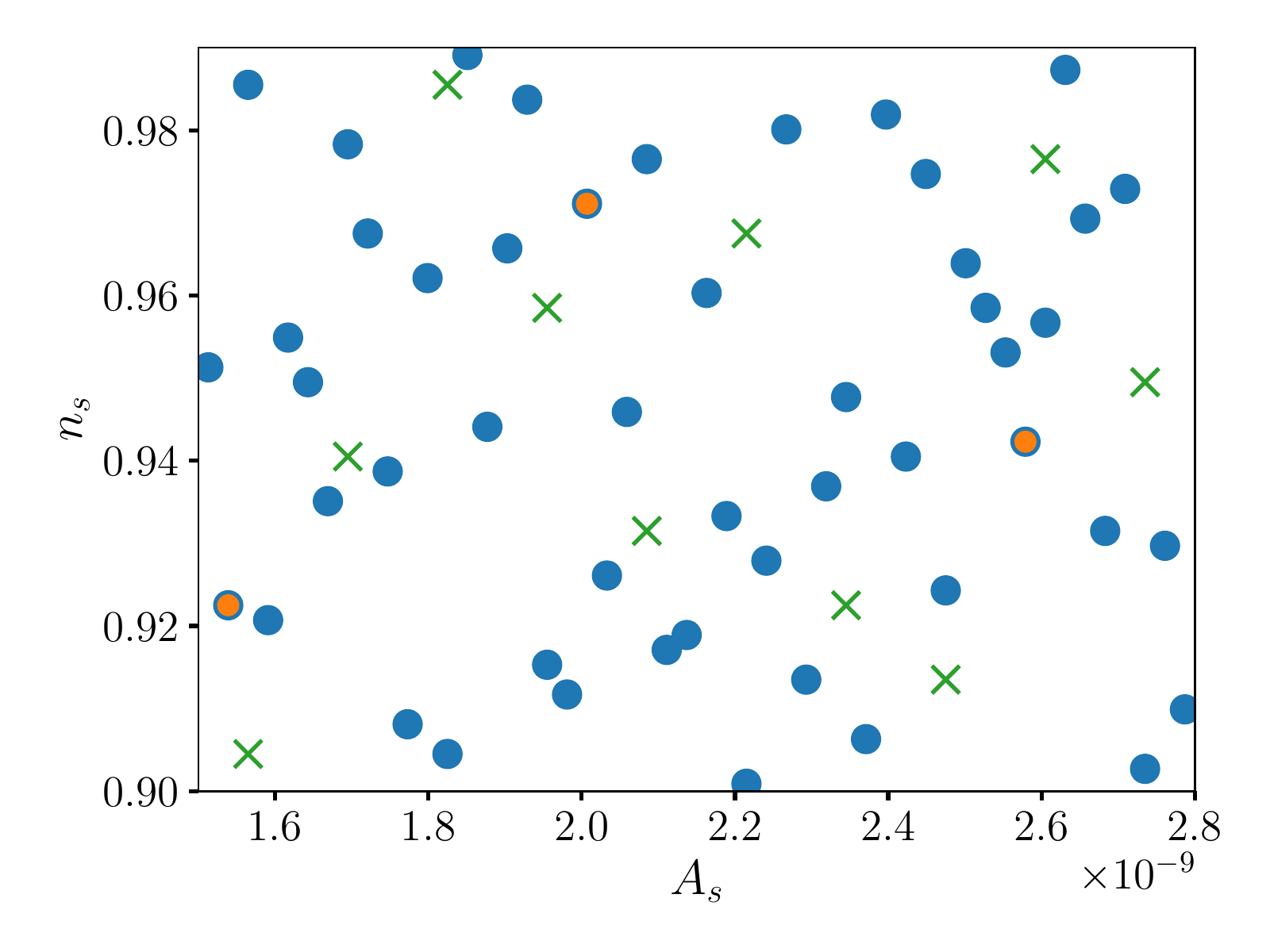}
   \caption{Two 2-D cross-sections of the 5-D samples of input parameters.
   The input parameters are designed with a nested structure, $\xvec_1 \subseteq \xvec_2$,
   between {\highres} and {\lowres}.
   \textbf{(Blue):}
   $\xvec_1$, 50 sampling points in {\lowres}.
   \textbf{(Orange):}
   $\xvec_2$, 3 sampling points in {\highres}.
   The selection of these $3$ points is chosen by the procedure described in Section~\ref{subsec:optimze_loss_low_fidelity},
   which minimizes the {\lowres} error in the low-fidelity only emulator.
   \textbf{(Green):} 10 points from the {\highres} testing set, which is a different Latin hypercube than $\xvec_1$.}
   \label{fig:sampling_nested_structure}
\end{figure*}

% [training outputs and testing outputs]
\begin{figure*}
   \includegraphics[width=\columnwidth]{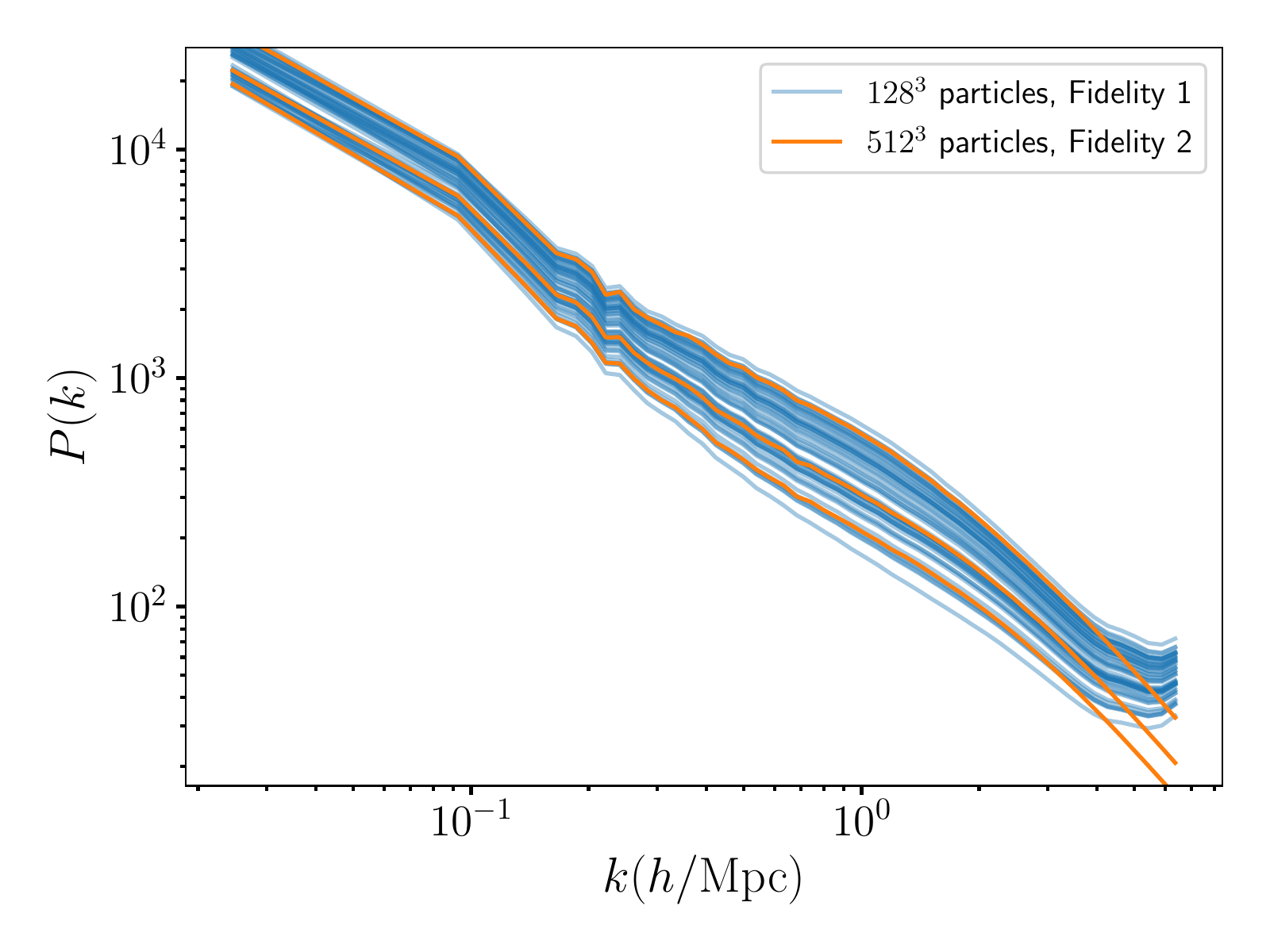}
   \includegraphics[width=\columnwidth]{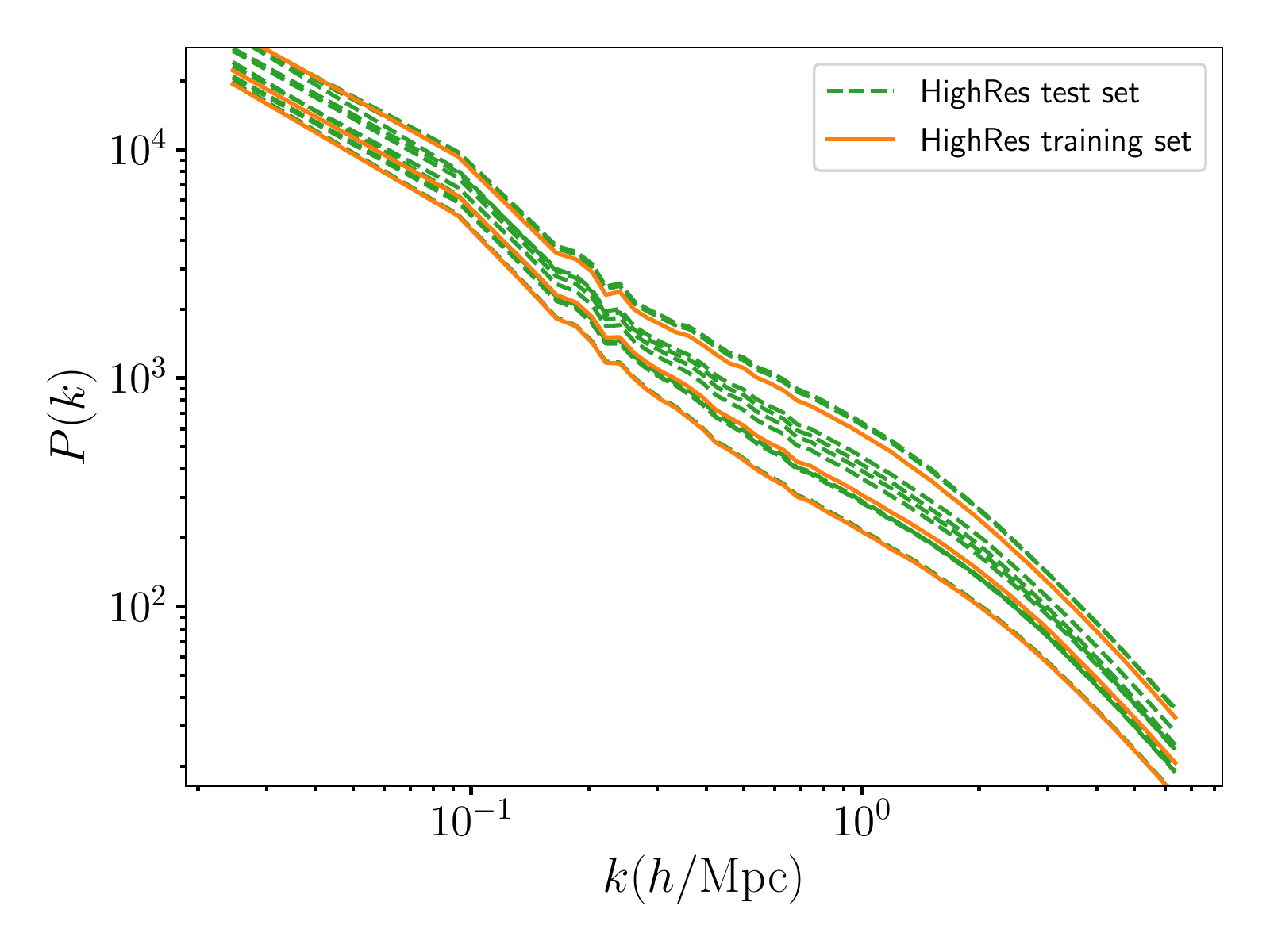}
   \caption{Training (left) and testing (right) data for the multi-fidelity emulator.
   \textbf{(Left):} $50$ low-fidelity training simulations (blue) and $3$ high-fidelity simulations (orange) used in a $50${\lowres}-$3${\highres} emulator.
   A {\highres} is a $512^3$ simulation and a {\lowres} is a $128^3$ simulation.
   Both {\highres} and {\lowres} are in a box with $256 \Mpch$ per side.
   The $50$ low-fidelity training simulations are drawn from a 5D Latin hypercube,
   $(h, \Omega_0, \Omega_b, A_s, n_s)$.
   The $3$ high-fidelity simulations are a subset of the low-fidelity simulation hypercube.
   \textbf{(Right):} $10$ high-fidelity test simulations (green dashed) and $3$ high-fidelity training simulations (orange).
   }
   \label{fig:training_testing_data}
\end{figure*}

% In terms of our design,
% we have two fidelity levels, $s = 2$, and we want our {\highres} input to be a subset of the {\lowres} input:
% \begin{equation*}
%    \xvec_{1} \subseteq \xvec_{2},
% \end{equation*}
% which means the input cosmological parameters from {\highres} are a subset of {\lowres} parameters. \spb{This subsection could be more focussed, I think. You already explained about a nested design in the last section so maybe we could just summarize the simulations run here?}

The proposed sampling scheme for training and testing is shown in Figure~\ref{fig:sampling_nested_structure}.
The corresponding output power spectra are shown in Figure~\ref{fig:training_testing_data}.
In Figure~\ref{fig:sampling_nested_structure},
the sampling is done using two different Latin hypercubes:
\begin{enumerate}
   \item Training simulations: a Latin hypercube with $50$ points. {\highres} points are a subset of {\lowres} points.
   \item Testing simulations: another Latin hypercube with $10$ points.
   \item We use the notation ``\xyemulator{$X$}{$Y$}'' to represent a multi-fidelity emulator trained on $X$ number of low resolution simulations and $Y$ number of high resolution simulations.
\end{enumerate}
The first hypercube with $50$ points ensures that we will have a nested experimental design.
The second hypercube is to ensure we will not test on the training simulations during the validation phase.
In practice, we found that the emulation accuracy roughly converged with $\sim 30$ {\lowres} points.

\subsection{Optimizing the loss of low-fidelity simulations}
\label{subsec:optimze_loss_low_fidelity}

For a multi-fidelity problem, we want to minimize the required high-fidelity training simulations to achieve a given accuracy.
We search for the optimal subset of {\lowres} points to simulate at {\highres} by picking the subset that would minimize the low fidelity training set's single-fidelity emulator errors. In our experiments with two fidelities, $s = 2$,
there are $\binom{n_1}{n_2}$ possible combinations for $\xvec_2$,
which are input parameters for the high-fidelity data, $\Data_2 = \{\xvec_2, \yvec_2\}$.
% For low-fidelity data $\Data_t$ with $t < s$, where $s$ is the highest fidelity,
% there will be $\binom{n_t}{n_s}$ possible combinations for $\Data_s$.

Retraining low-fidelity only emulators on all possible subsets of the low fidelity grid is computationally intensive.
For example, selecting two samples out of $50$ points means that we have to train $\binom{50}{2} = 1\,225$ low-fidelity emulators.
To save computational cost, we employed a greedy optimization strategy.
Instead of exploring all possible subsets, we grew the subset one point at a time, fixing the previously chosen points.
As a further optimisation, we used the same set of kernel hyperparameters for all $k$ bins.

Consider $\Ssubset$, a potential $\Data_2$ with $\xvec_2 \subset \xvec_1$.
We train a low-fidelity only emulator based on Eq~\ref{eq:mle} using the $n_2$ low-fidelity points in $\Ssubset$ and get a {\gp}:
\begin{equation}
   p(f_{*} \mid \Ssubset, x_*) = \normal(f_{*} \mid \mu^{(i)}_*(x_*), \sigma_{*}^{(i)}(x_*)^2),
   \label{eq:subset_sf_gp}
\end{equation}
which is the posterior as described in Eq~\ref{eq:sf_posterior_pred}.

With the trained low-fidelity only emulator in Eq~\ref{eq:subset_sf_gp},
we can test this single-fidelity emulator's performance by predicting the rest of the data in the low-fidelity Latin hypercube.
To evaluate the accuracy, we compute the mean squared error by averaging over the test data:
\begin{equation}
   \mathrm{MSE} = \mathbb{E}[ ( y_* - \mu_*^{(i)}(x_*) )^2 ],
\end{equation}
where $\{(x_*, y_*)\}$ are the low-fidelity data pairs from the rest of the Latin hypercube,
\begin{equation}
   \{(x_*, y_*)\} \in \{ \Data_1 - \Ssubset \}\,.
\end{equation}
This simply means that we test the single-fidelity emulator on the available data not included in the training subset.

Suppose we repeat the training of single-fidelity emulators until we train all possible subsets in the low-fidelity hypercube.
We will now have $\binom{n_1}{n_2}$ trained single-fidelity emulators.
Each single-fidelity emulator will provide a mean squared error,
which is the test error that the emulator generates against the low-fidelity hypercube test data.
To select the optimally trained emulator, we compute
\begin{equation}
   \Ssubset^* = \mathrm{arg\,min}_{\Ssubset^*} (\mathbb{E}[ ( y_* - \mu_*^{(i)}(x_*) )^2 ]),
   \label{eq:low_res_argmin}
\end{equation}
where we find the subset $\Ssubset^*$ which minimizes the mean squared errors on the test set.
We use $\Ssubset^*$ as our high-fidelity training set $\Data_{2}$
under the nested experimental design.
To be explicit:
\begin{equation}
   \xvec_{2} = \xvec_{\Ssubset^*} \subset \xvec_{1},
\end{equation}
where $\xvec_{2}$ are the selected high-fidelity input points,
$\xvec_{\Ssubset^*}$ are the input points from the selected subset $\Ssubset^*$ (which minimize the low-fidelity emulator mean squared error), and $\xvec_{1}$ are the low-fidelity input points.

% write out the underlying assumption
This strategy assumes that the effect of a sampling scheme on a low-fidelity emulator is the same as that on a corresponding multi-fidelity emulator.
For example, suppose $\Delta \Omega_b$ is crucial for learning how the low-fidelity power spectrum $y_{1}$ changes for inputs $x_{1}$. In that case, we expect that information about $\Delta \Omega_b$ can also effectively change the high-fidelity spectrum
$y_{2}$.

The above assumption could be violated if the power spectra at small scales, which are not included in the low-fidelity data, behave very differently from those at large scales.
This could happen if the smoothness length scale acts very differently between low-fidelity and high-fidelity data for a given input dimension.
For example, imagine that a parameter, $\theta$, has a small effect on the outcomes of low-fidelity simulations, but a large effect on the outcomes of high-fidelity simulations.

% [figure] an example figure of correlation between low-fidelity emulator error and high-fidelity emulator error
\begin{figure}
   \includegraphics[width=\columnwidth]{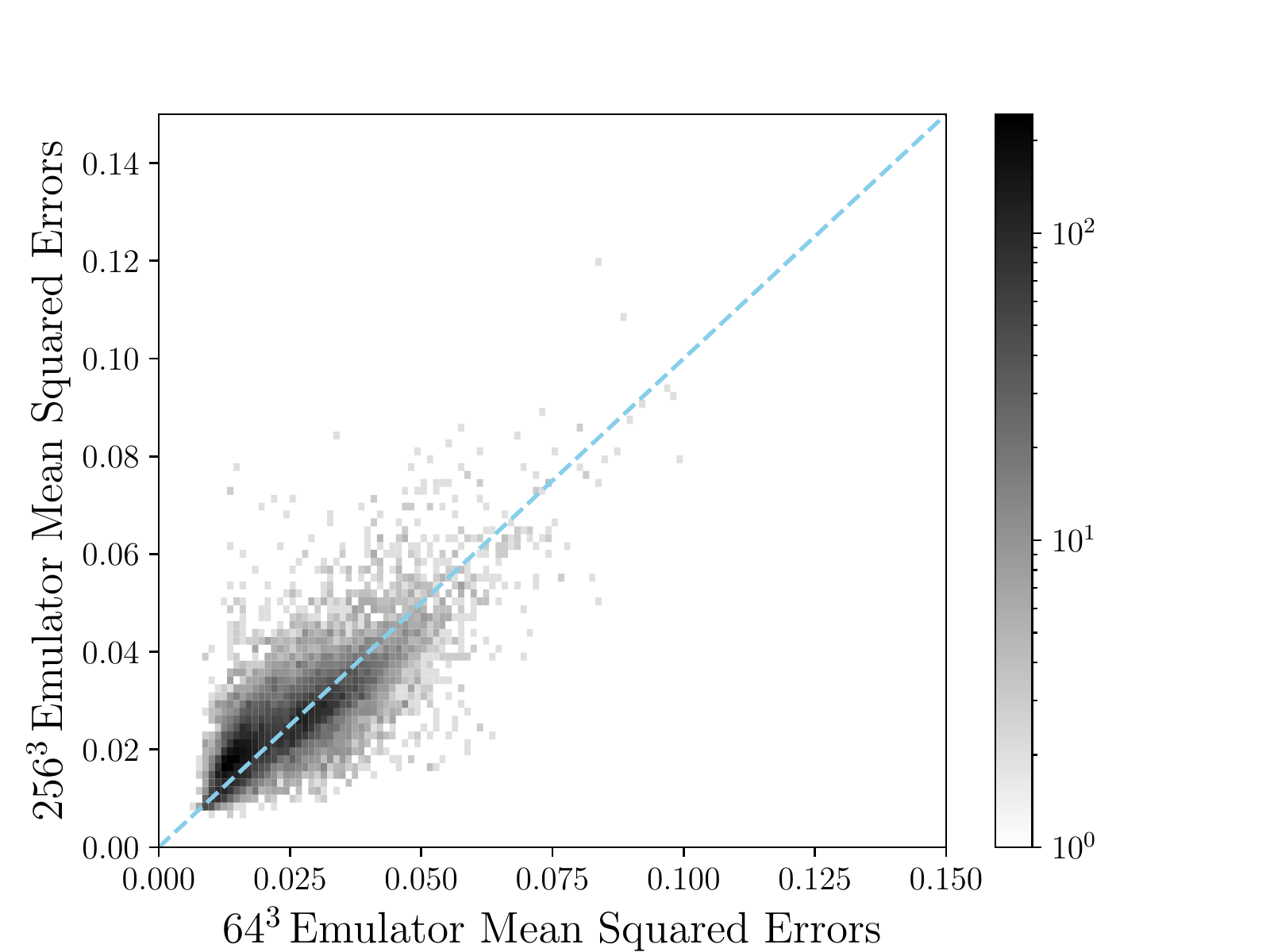}
   \caption{Emulator mean squared errors evaluated from $64^3$ emulators and $256^3$ emulators.
   We compute all subsets of $3$ samples from a $50$ samples Latin hypercube, $\binom{50}{3} = 19\,600$ subsets in total.
   Colorbar is in log scale. The blue dashed line represents a perfect linear relationship.}
   \label{fig:loss_50_3}
\end{figure}

Figure~\ref{fig:loss_50_3} shows the mean squared errors computed from $64^3$ single-fidelity emulators and $256^3$ single-fidelity emulators.
First, note that the selection of the training simulations affects the emulator accuracy.
Second, the low-fidelity emulator errors are correlated with their higher fidelity counterparts.
This suggests that a low-fidelity emulator can serve as a guide for placing high-fidelity training simulations.
The HR parameter choices used in Section~\ref{sec:results} were selected with an earlier version of our model using $64^3$ particle simulations.
We checked that using either $64^3$ or $128^3$ for selection gave almost the same emulation accuracy for a non-linear \xyemulator{50}{3}, though one of the selected samples is different.

% [todo] plot MF test errors as a function of lowRes selection test errors
% need to re-run all possible MF emulators with the linear theory GP mean prior update.
In practice, we find the procedure above can prevent us from selecting the {\highres} combination that will give us the worst multi-fidelity emulation result.
Although we have tested that our procedure works for the matter power spectrum,
we would suggest that when emulating a new summary statistic (e.g., the halo mass function), the reader investigates the effectiveness of this method using small test cases.
We may in future work investigate using Bayesian optimization \cite[e.g.,][]{Forrester:2007,Lam:2015,MISO:2016} to select the optimal {\highres} samples for multi-fidelity training.

\section{Results}
\label{sec:results}

This section shows the interpolation accuracy of multi-fidelity methods and compares our multi-fidelity emulators to single-fidelity emulators.
Section~\ref{subsec:emu_errors} compares test set emulator errors for the linear multi-fidelity emulator (AR1) and non-linear multi-fidelity emulator (NARGP).
Section~\ref{subsec:compare} compares a multi-fidelity emulator to two kinds of single-fidelity emulators: high-fidelity only and low-fidelity only.
We also compare the emulator accuracy as a function of core hours for both multi-fidelity emulators and single-fidelity emulators.

To test how much a multi-fidelity emulator can improve with more training simulations, Section~\ref{subsec:vary_train_simulations} shows the emulator errors with more {\lowres} or {\highres} training simulations.
Finally, Section~\ref{subsec:check_other_settings} checks the performance of the multi-fidelity method for other emulation settings.

\subsection{Comparison of Linear and Non-Linear Emulators}
\label{subsec:emu_errors}

\begin{figure}
   \includegraphics[width=\columnwidth]{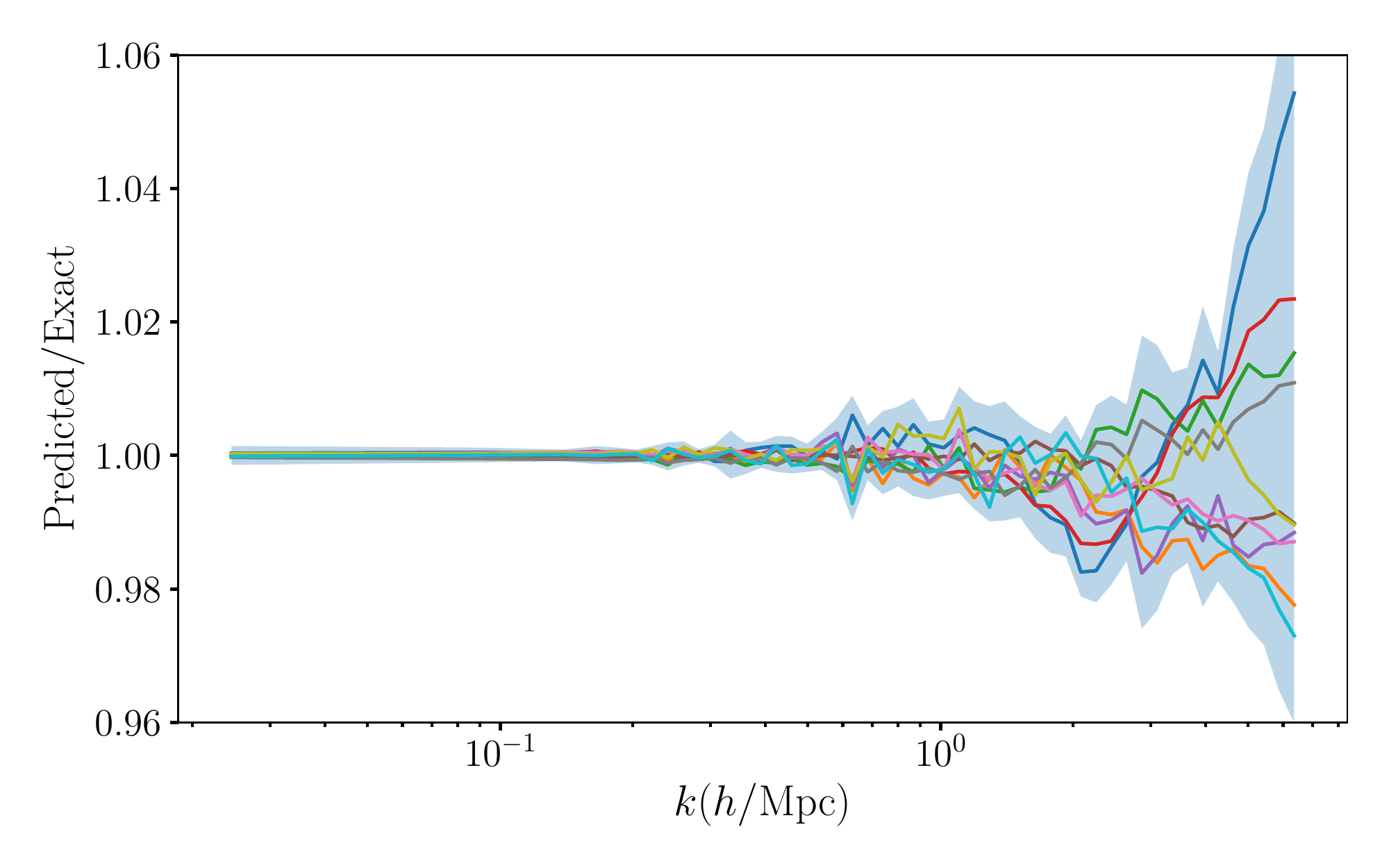}
   \caption{Predicted divided by exact power spectrum from a \xyemulator{50}{3} using a linear multi-fidelity method (AR1).
   Different colours correspond to $10$ test simulations spanning a $5$-D Latin hypercube.
   The shaded area indicates the worst-case $1-\sigma$ emulator uncertainty.
   There is one test simulation driving the larger error compared to the non-linear one in Figure~\ref{fig:pred_exact_3_nonlinear}.}
   \label{fig:pred_exact_3}
\end{figure}

\begin{figure}
   \includegraphics[width=\columnwidth]{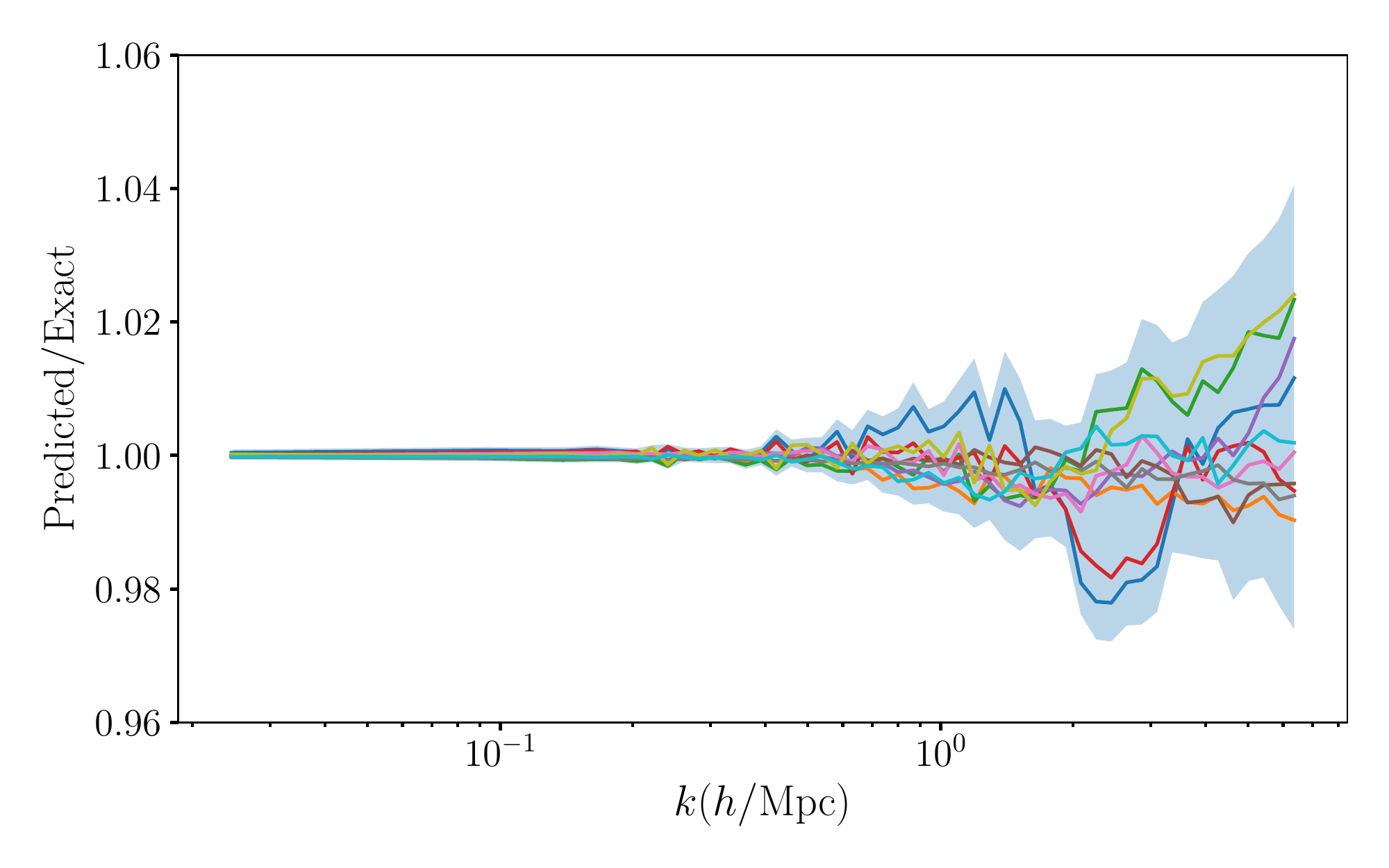}
   \caption{Predicted divided by exact power spectrum from a \xyemulator{50}{3} using a non-linear multi-fidelity method (NARGP).
   Different colours correspond to $10$ test simulations spanning a $5$-D Latin hypercube.
   The shaded area indicates the worst-case $1-\sigma$ emulator uncertainty.
   Note that the y-scale in this plot is the same as Figure~\ref{fig:pred_exact_3}.}
   \label{fig:pred_exact_3_nonlinear}
\end{figure}

Figure~\ref{fig:pred_exact_3} and Figure~\ref{fig:pred_exact_3_nonlinear} show the predicted power spectrum divided by the exact power spectrum for simulations in the testing set.
Both emulators, linear (AR1) and non-linear (NARGP), are trained with $50$ low-fidelity simulations and $3$ high-fidelity simulations.
We will call these emulators ``\xyemulator{50}{3}s'' for simplicity.
A non-linear (linear) multi-fidelity emulator requires at least $3$ (2) {\highres} simulations for training and has $\lesssim 2\%$ ($\lesssim 5\%$) worst-case accuracy per $k$ bin.
For a linear multi-fidelity emulator, the minimum required number of {\highres} simulations is $2$, reflecting the lower number of hyperparameters in the kernel.

\begin{figure}
   \includegraphics[width=\columnwidth]{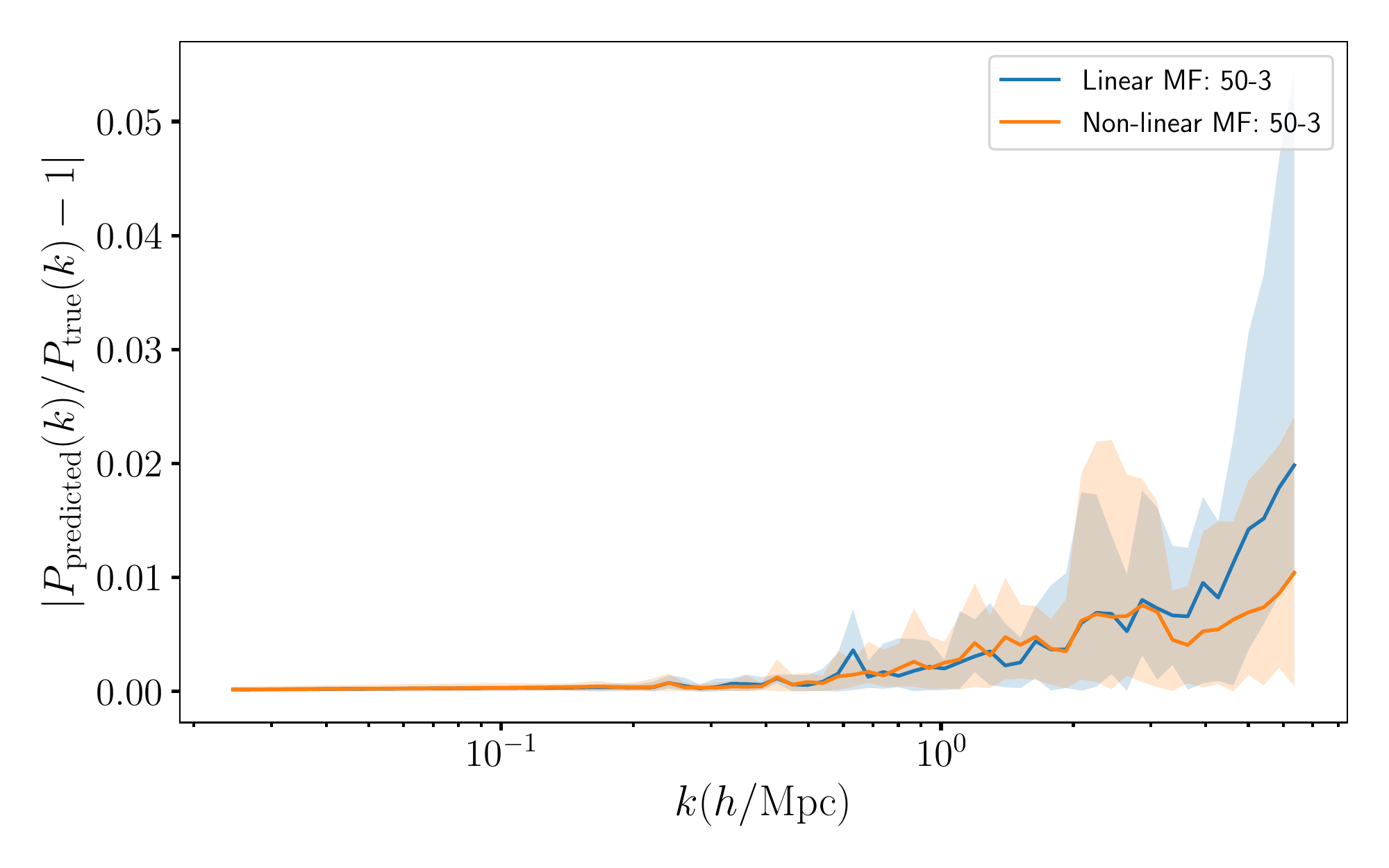}
   \caption{Relative emulator errors from a \xyemulator{50}{3} using
   linear multi-fidelity (blue) and non-linear multi-fidelity (orange).
   % We also include a linear \xyemulator{50}{2} (green).
   Solid lines represent the average error from test simulations, $\frac{1}{10} \sum_{i=1}^{10} | \frac{P_\mathrm{pred}{_{,i}}}{P_\mathrm{true}} - 1 |$.
   Shaded areas show the maximum and minimum test errors.
   }
   \label{fig:linear_nonlinear_errors_3hr}
\end{figure}

Figure~\ref{fig:linear_nonlinear_errors_3hr} shows a comparison between a linear multi-fidelity emulator and a non-linear multi-fidelity emulator in relative emulator error.
We include linear and non-linear \xyemulator{50}{3}s.
We define the relative emulator error:
\begin{equation}
   \mathrm{Emulator\,Error} = \left| \frac{P_\mathrm{pred}}{P_\mathrm{true}} - 1 \right|.
   \label{eq:emulator_error}
\end{equation}
$P_\mathrm{pred}$ is the predicted power spectrum from the multi-fidelity emulator, and $P_\mathrm{true}$ is the power spectrum from the high-fidelity test simulation.

Figure~\ref{fig:linear_nonlinear_errors_3hr} shows that the linear \xyemulator{50}{3} predicts an average error $< 1\%$ per $k$ bin for $k \leq 4 \,\hMpc$ and $< 2\%$ per $k$ bin for $4 < k \leq 6.4 \,\hMpc$.
The non-linear multi-fidelity emulator predicts an average error $\lesssim 1\%$ per $k$ bin, which implies we only need $3$ {\highres} to achieve a percent-level accurate emulator using the non-linear multi-fidelity method.
At $k \leq 3 \hMpc$, both emulators predict mostly the same accuracy,
but the non-linear one performs better at smaller scales $k > 3 \hMpc$.

% main figure on emulator error with 3 HR (linear/nonlinear)

% Referee's suggestion - do not mention 50-2 emulator
%
% Figure~\ref{fig:linear_nonlinear_errors_3hr} the comparison between linear multi-fidelity and non-linear multi-fidelity in terms of relative emulator errors.
We found that the non-linear multi-fidelity emulator outperforms the linear one in all aspects.
For simplicity, we will only show the non-linear multi-fidelity models in the following sections, but we note that a linear multi-fidelity model is still useful when only two {\highres} simulations are available.
% A linear multi-fidelity has fewer hyperparameters than a non-linear multi-fidelity,
% so it is trainable with only $2$ {\highres}.
We also found that, for the linear model, changing from \xyemulator{50}{3} to \xyemulator{50}{2} only slightly degrades the overall accuracy.
% Nevertheless, to make consistent comparisons, we will only compare emulators with at least $3$ {\highres} from now on.

\subsection{Comparison to single-fidelity emulators}
\label{subsec:compare}

% main comparison plots
\subsubsection{Comparison to high-fidelity only emulators}

% mf 3-11; nonlinear model
\begin{figure}
   \includegraphics[width=\columnwidth]{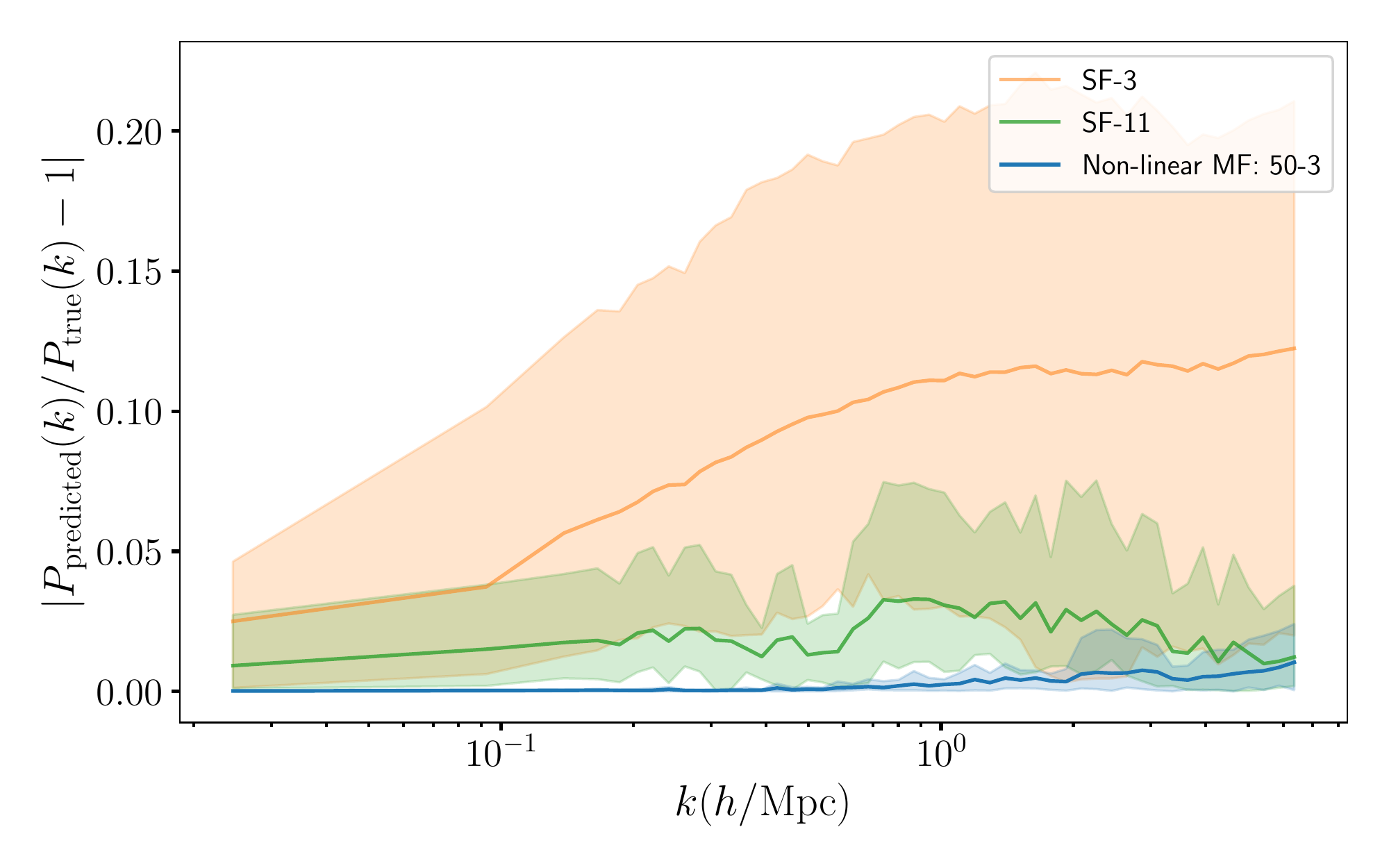}
   \caption{\textbf{Non-linear multi-fidelity emulator (blue)} with $50$ {\lowres} and $3$ {\highres} simulations,
   compared to \textbf{single-fidelity emulators} with $3$ {\highres} (orange) and with $11$ {\highres} (green).
   Shaded area indicates the maximum and minimum emulation errors.
   The computational cost for a \xyemulator{50}{3} $\simeq 9\,000$ core hours
   while the single-fidelity emulator with $11$ {\highres} requires $\simeq 25\,000$ core hours.
   However, a \xyemulator{50}{3} still outperforms an 11\,{\highres} emulator.}
   \label{fig:single_multi_comparison_3_11_nonlinear}
\end{figure}

Figure~\ref{fig:single_multi_comparison_3_11_nonlinear} shows a comparison between a non-linear \xyemulator{50}{3} and high-fidelity only emulators.
The high-fidelity only emulators are single-fidelity emulators trained solely on {\highres} simulations.
The non-linear multi-fidelity emulator outperforms the single-fidelity emulator with $11$ {\highres} at all $k$ modes.
It also predicts a worst-case error smaller than the worst-case error from the 11\,{\highres} single-fidelity emulator.
At $k \leq 2 \hMpc$,
the multi-fidelity emulator performs much better than the single-fidelity emulators.
Since {\lowres} simulations can predict accurate power spectrum at large scales $k \leq 2 \hMpc$,
we expect a single-fidelity emulator requires $\sim 50$ {\highres} to compete with the \xyemulator{50}{3} on large scales.
A {\highres} is $\simeq 64$ times more expensive than a {\lowres},
thus the core time for a \xyemulator{50}{3} is $\simeq 4$ {\highres}.
The non-linear multi-fidelity outperforms a single-fidelity 11\,{\highres} emulator with $\simeq 3$ times lower computational cost.

% Referee suggested - remove this figure
%
% % rate of reduction
% \begin{figure}
%    \includegraphics[width=\columnwidth]{images/mf_sf_error_reduction_per_bin_3-3nonlinear.pdf}
%    \caption{Error reduction rate from single-fidelity to multi-fidelity emulator.
%    We fix the number of {\highres} simulations to $3$, and calculate the reduction of average emulator errors from a linear \xyemulator{50}{3} (orange) and a non-linear \xyemulator{50}{3} (green). A larger reduction rate implies a better multi-fidelity emulator at a given scale.}
%    \label{fig:mf_sf_error_reduction}
% \end{figure}

% Figure~\ref{fig:mf_sf_error_reduction} depicts the error reduction rate between a \xyemulator{50}{3} and a 3-{\highres} single-fidelity emulator.
The error reduction rate is the relative error of a single-fidelity emulator divided by the error of a multi-fidelity emulator.
Both linear and non-linear \xyemulator{50}{3} show an error reduction rate of $\simeq 100$ for $k \leq 0.5 \hMpc$, $ \simeq 100$ times better than the single-fidelity counterpart using $3$ {\highres}.
At smaller scales $ k > 3 \hMpc$,
the multi-fidelity emulators are $\simeq 20$ times (non-linear), and $\simeq 10$ times (linear) better than their single-fidelity counterpart.

\subsubsection{Comparison to low-fidelity only emulators}
\label{subsec:lowfideliy_emulator}

\begin{figure}
   \includegraphics[width=\columnwidth]{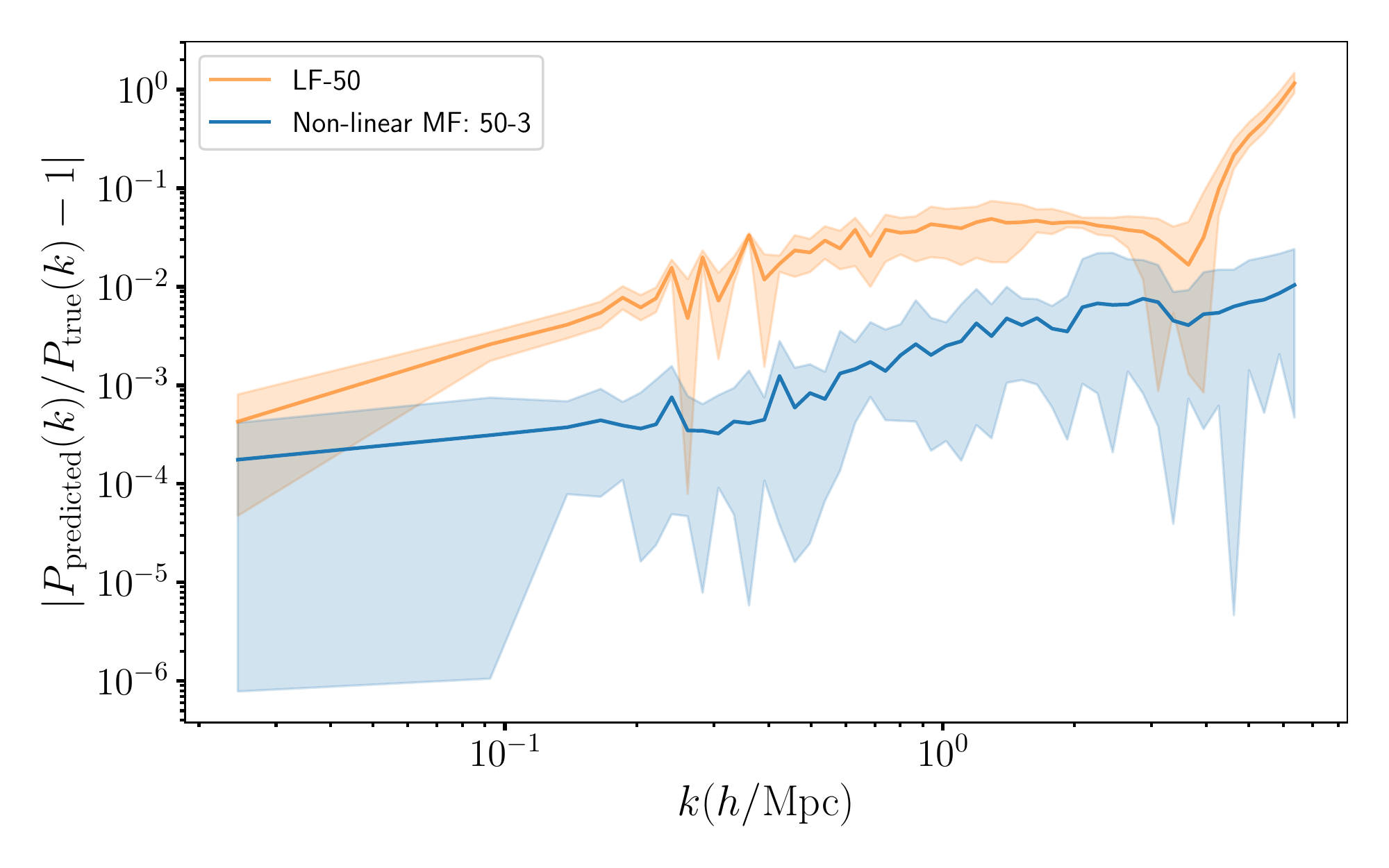}
   \caption{Relative emulator errors between a $50$ low-fidelity emulator and a non-linear \xyemulator{50}{3}.
   Errors are evaluated on $10$ {\highres} simulations.
   Shaded area indicates the maximum and minimum errors.
   Note that the y-axis is in $\log_{10}$ scale.
   }
   \label{fig:lowres_50}
\end{figure}

Figure~\ref{fig:lowres_50} shows a single-fidelity emulator trained on $50$ {\lowres} simulations, compared to a non-linear \xyemulator{50}{3}.
Figure~\ref{fig:lowres_50} demonstrates how multi-fidelity modelling improves the emulator accuracy at each $k$ scale.
At $k \lesssim 3 \hMpc$,
multi-fidelity modelling uses $3$ {\highres} to correct the resolution and reduce the average emulator error from $\lesssim 5\%$ to $\leq 1 \%$.
A low-fidelity emulator predicts a biased power spectrum beyond $k = 3 \hMpc$.
However, the multi-fidelity method can moderately correct the bias and reduce the error to $\lesssim 1 \%$.
Again, the multi-fidelity technique can use a few {\highres} simulations to calibrate the resolution difference.

\subsubsection{Core hours versus emulator errors}
\label{subsubsec:core_hours}

\begin{figure}
   \includegraphics[width=\columnwidth]{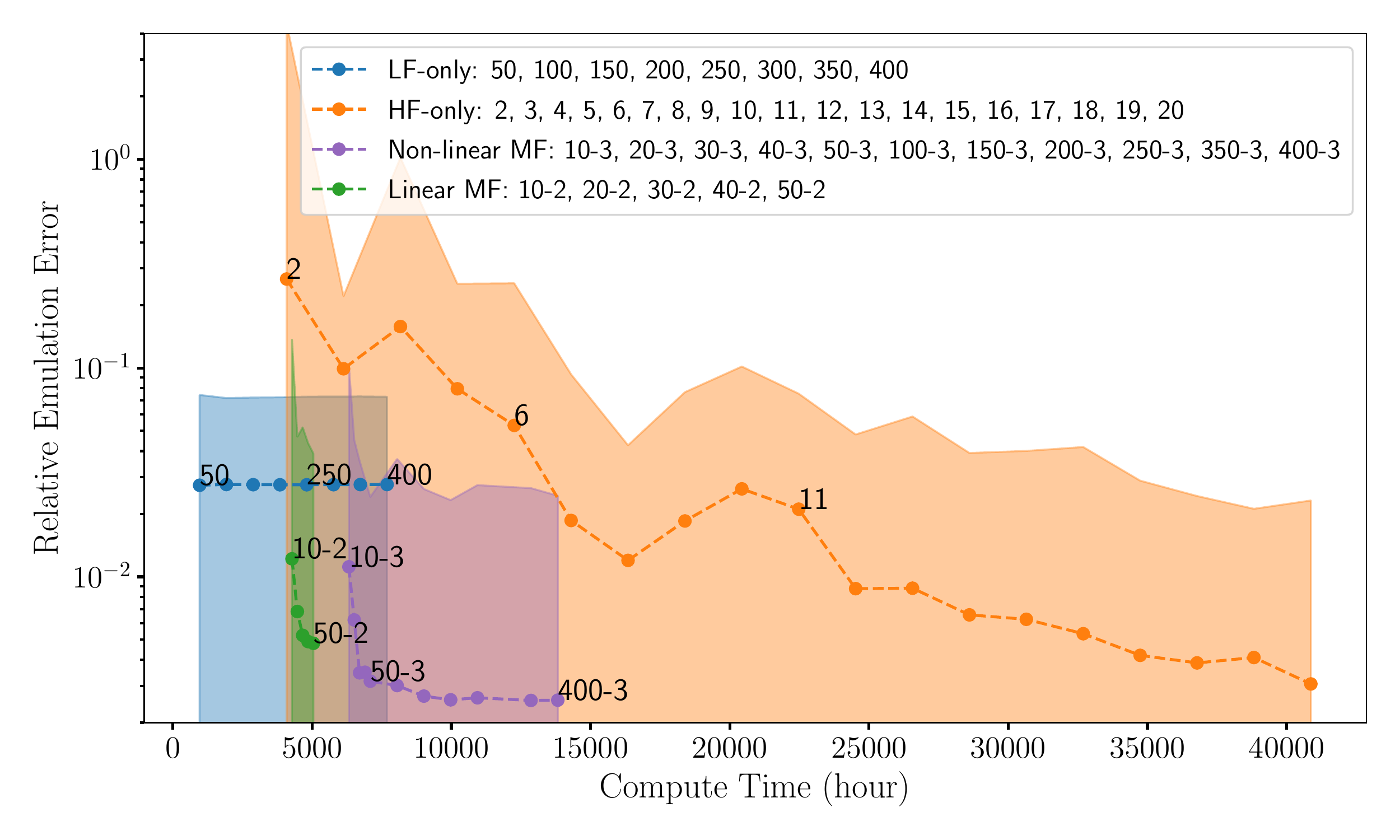}
   \caption{Core hours for running the training simulations versus emulation errors for high-fidelity only emulators (orange) and low-fidelity only emulators (blue), linear multi-fidelity emulators (AR1) with $2$ {\highres} (green),
   and non-linear multi-fidelity emulators (NARGP) with $3$ {\highres} (purple).
   The numbers in the labels indicate the number of training simulations used in the emulator.
%    For example, the numbers after LF-only emulator mean how many low-resolution simulations used in the emulator.
   For multi-fidelity emulators, $X$-$Y$, $X$ is the number of low-resolution and $Y$ is the number of high-resolution training simulations.
   The dots show the average errors.
   The upper shaded areas show the maximum emulator errors among 10 test simulations.
   The {\lowres} samples beyond $100$ are drawn from a separate Latin hypercube with $400$ samples.
   For LF-only emulators, we only calculate the relative errors for $k \leq 3$.}
   \label{fig:compute_time_accuracy}
\end{figure}

Figure~\ref{fig:compute_time_accuracy} shows the average relative emulator error as a function of core hours for performing the training simulations.
The emulator errors shown in Figure~\ref{fig:compute_time_accuracy}
are averaged over all $k$ modes, so each emulator corresponds to a single point in the plot.
An ideal emulator will be on the left bottom corner, implying both low cost and high accuracy.
The slope of a given emulator in the plot indicates how easily we can improve the emulator with more training data.
A steeper (more negative) slope means we can increase the emulator accuracy with a lower cost.

We notice three types of emulators are clustered in separate regions in the plot.
The low-fidelity only emulator (LF-only) has the lowest cost and shows no noticeable improvement from increasing training simulations from $50$ to $400$ {\lowres}.
The high-fidelity only emulator (HF-only) shows an accuracy improvement with more {\highres} simulations from $3$ {\highres} to $11$ {\highres}.
However, performing one {\highres} requires $\sim 2\,000$ core hours, making the HF-only emulator much more expensive than the other two emulators in the plot.

In Figure~\ref{fig:compute_time_accuracy}, the non-linear multi-fidelity emulator (NARGP) shows a compute time similar to $3$ {\highres} simulations but has better accuracy than the HF-only emulator.
It also presents a steeper slope than the HF-only emulator, indicating we can efficiently increase the accuracy using low-cost {\lowres} simulations.
From \xyemulator{10}{3} to \xyemulator{50}{3}, it shows that we can decrease the error from $\sim 0.02$ to $\sim 0.003$ using an additional $\sim 800$ core hours.
From \xyemulator{50}{3} to \xyemulator{400}{3}, we also see a mild decrease of error but not as steep as $10${\lowres}-$3${\highres} to $50${\lowres}-$3${\highres}.

We also include the linear model (AR1) to demonstrate the performance of the multi-fidelity method when there are only $2$ {\highres} available.
The linear model also shows a steep improvement slope from $10${\lowres}-$2${\highres} to $50${\lowres}-$2${\highres}.
However, we notice that the linear model with $2$ {\highres} is slightly worse than the non-linear one with $3$ {\highres}.

Figure~\ref{fig:compute_time_accuracy} demonstrates that a multi-fidelity emulator can provide good accuracy with a much lower cost than HF-only emulators. It also points out that we can efficiently improve the accuracy of a multi-fidelity emulator using cheap low-fidelity simulations.

\subsection{Varying the number of training simulations}
\label{subsec:vary_train_simulations}

\subsubsection{Effects of more low-resolution training simulations}
\label{subsec:adding_lowres}

% figure emulator error per k bins with increasing number of lowres
\begin{figure}
   \includegraphics[width=\columnwidth]{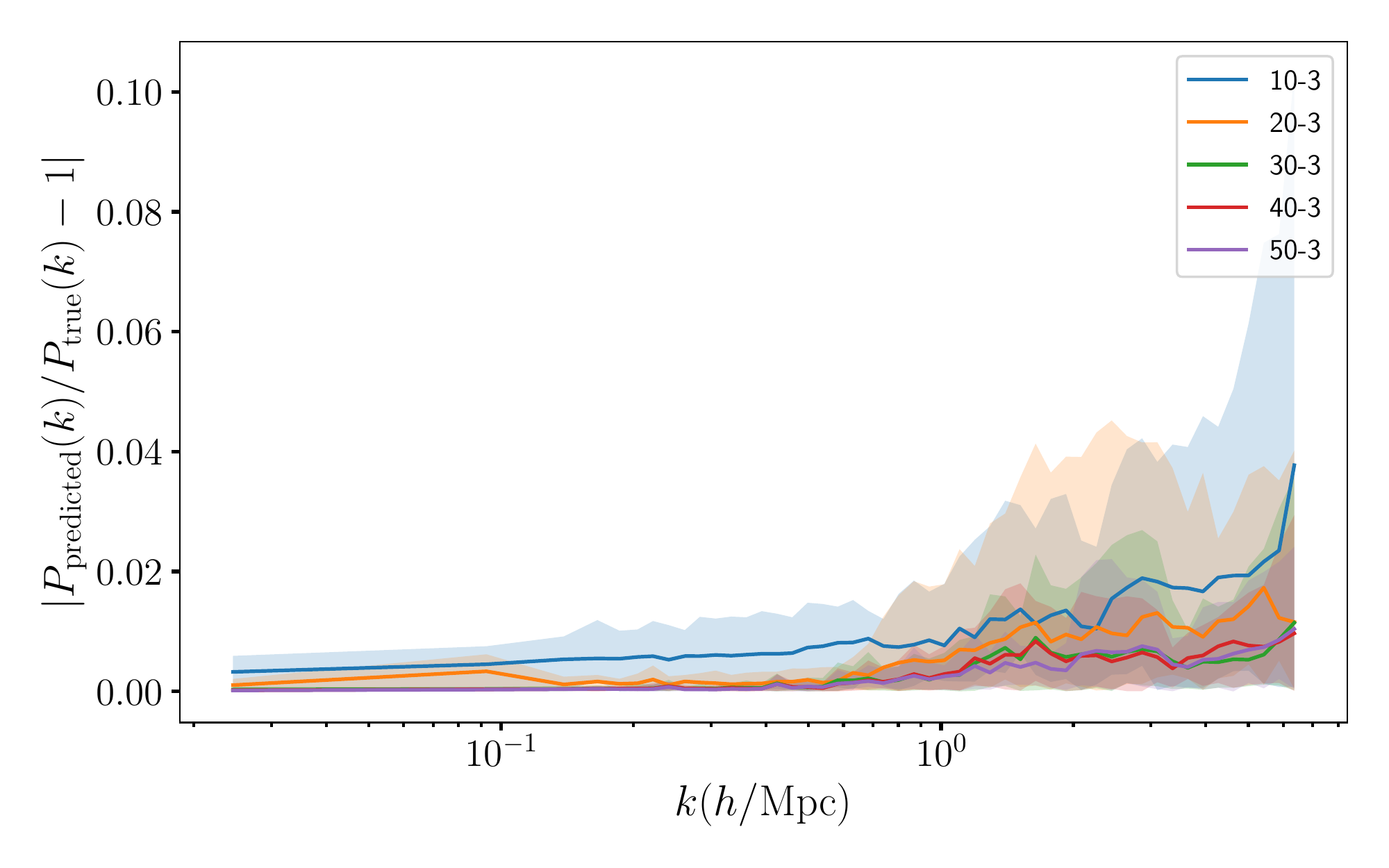}
   \caption{
      Relative emulator error of non-linear \xyemulator{$N$}{3} colour coded with different number of {\lowres} training simulations, with $N \in \{10, 20, 30, 40, 50\}$.
      The same as Figure~\ref{fig:linear_nonlinear_errors_3hr},
      solid lines represent the average error from test simulations, $\frac{1}{10} \sum_{i=1}^{10} | \frac{P_\mathrm{pred}{_{,i}}}{P_\mathrm{true}} - 1 |$, and
      shaded areas show the maximum and minimum test errors.
   }
   \label{fig:comparison_increase_lowres_per_k}
\end{figure}

The benefit of using a multi-fidelity emulator is that we can improve the emulator accuracy using extra low-fidelity simulations.
Figure~\ref{fig:comparison_increase_lowres_per_k} shows the emulator error colour coded by the number of {\lowres} training simulations.
With more {\lowres} training data, the emulator performance improves at both large and small scales.
We only show the non-linear emulator here for simplicity,
but we observe a similar trend in the linear emulator.
For $N${\lowres}-$3${\highres} with $N \in \{10, 20, 30, 40, 50\}$ emulators,
the last $k$ bin gives $3.77 \%$, $1.16\%$ , $1.15 \%$ , $0.97 \%$, and $1.04 \%$ emulator errors,
indicating an increase of accuracy with more {\lowres} training simulations.
Dividing the errors into large and small scales at $k = 1 \hMpc$,
the average emulator errors are $0.65 \%$, $0.22 \%$, $0.10 \%$, $0.09 \%$, and $0.09 \%$ for $k \leq 1 \hMpc$
and $1.60 \%$, $1.04 \%$, $0.60 \%$, $0.61 \%$, and $0.56 \%$ for $k > 1 \hMpc$.
The decrease in error is nearly saturated with $\sim 40$ {\lowres} simulations.

\subsubsection{Effects of more high-resolution training simulations}
\label{subsec:more_highres}

% nonlinear 3 HR - 7 HR
\begin{figure}
   \includegraphics[width=\columnwidth]{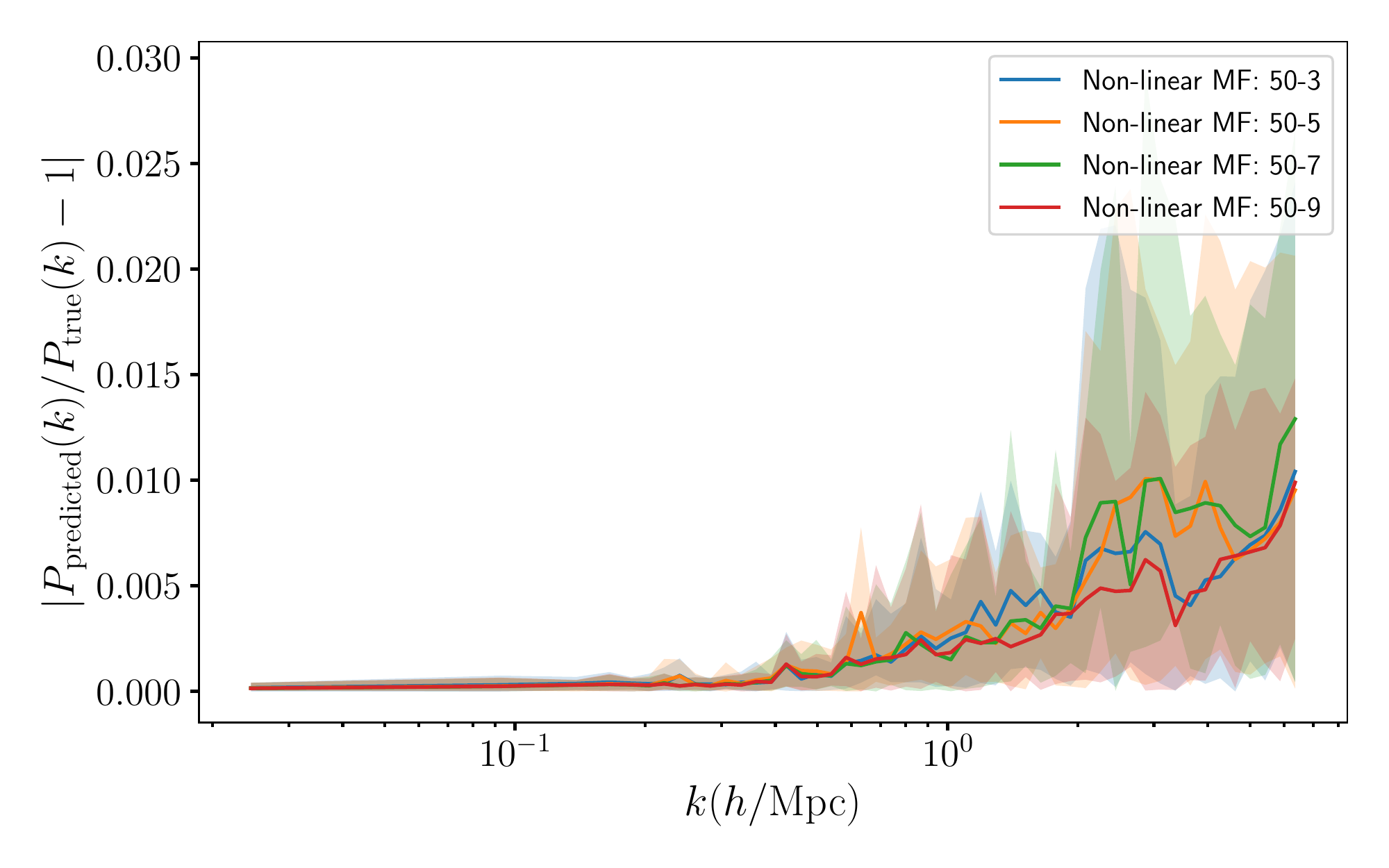}
   \caption{Relative emulator errors from non-linear \xyemulator{50}{$N$} with $N = 3$ (blue), $N = 5$ (orange), $N = 7$ (green), and $N = 9$ (red) {\highres} training simulations.
   Solid lines are the average test errors.
   Shaded areas show the maximum and minimum test errors.}
   \label{fig:comparison_nonlinear_3_5_7}
\end{figure}

In Figure~\ref{fig:comparison_nonlinear_3_5_7}, we add more {\highres} training simulations to our multi-fidelity emulator.
The \xyemulator{50}{N} with $N\in\{3, 5, 7, 9\}$ shows no improvement in average error with more {\highres}, although the worst case error improves noticeably for the \xyemulator{50}{9}.
One reason may be stochasticity in the training set due to simulation modelling error, which is around $1\%$, and limits the prediction accuracy.
In particular, {\mpgadget} simulations with $512^3$ particles may not be fully converged on small scales, and this limits the emulator's learning.
Another possibility is that the prior from $50$ low-fidelity simulations may be too hard to overcome with only $9$ {\highres} simulations.

To improve multi-fidelity emulator accuracy further, one could build a more complicated model than the one proposed in this paper.
The improvement from the linear to the non-linear model shows that different decisions about the scaling factor $\rho$ could better predict the non-linear structure.
However, those complicated models will require more high-fidelity training simulations.
We will leave more complex modelling structures to future work.

\subsection{Effect of other emulation parameters}
\label{subsec:check_other_settings}

\subsubsection{The resolution of low-fidelity simulations}
\label{subsec:change_lowres}

\begin{figure}
   \includegraphics[width=\columnwidth]{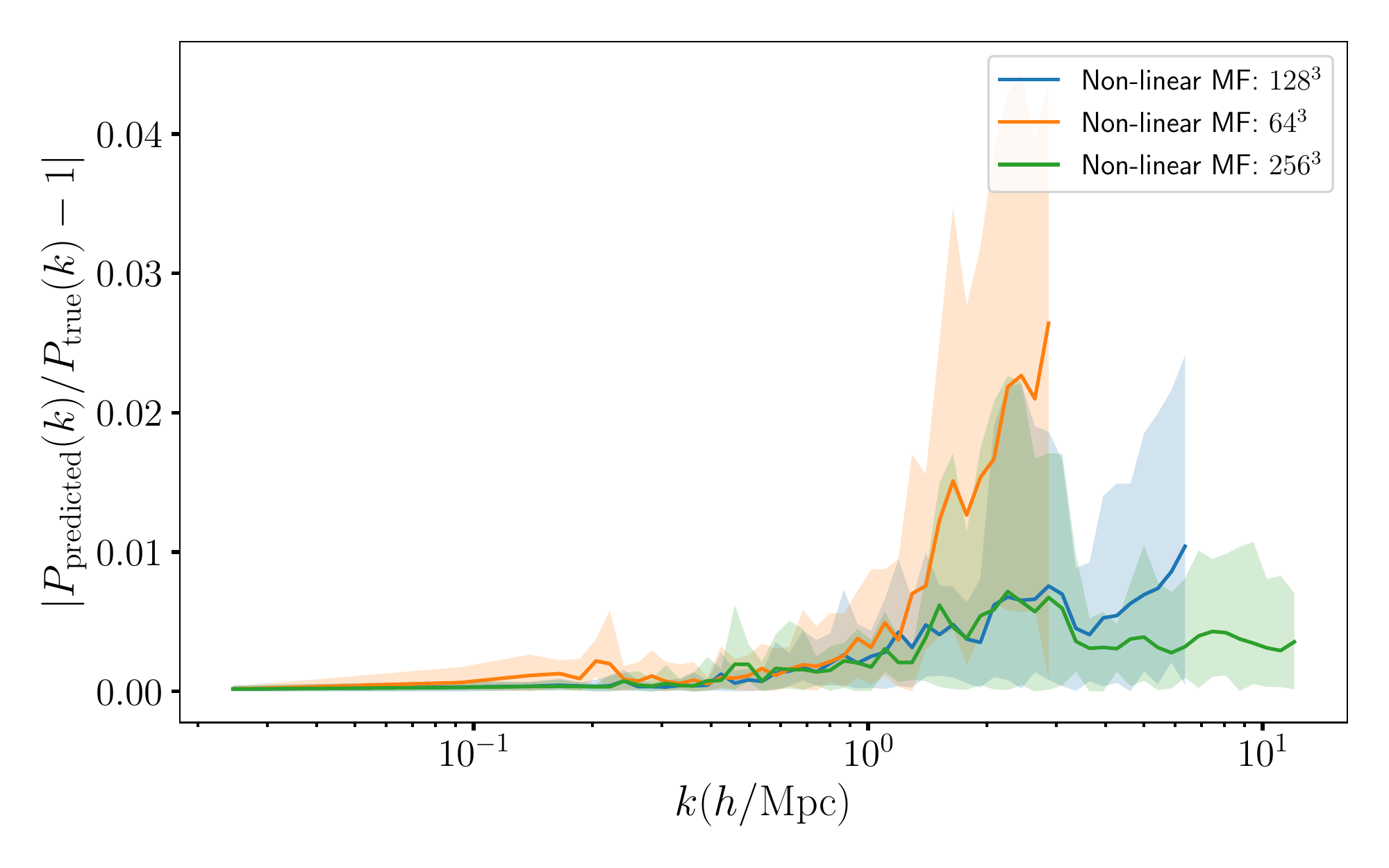}
   \caption{Relative emulator errors for \xyemulator{50}{3} emulators using different qualities of {\lowres} simulations.
   \textbf{(Blue)}: using $128^3$ simulations as low-fidelity training simulations.
   \textbf{(Orange)}: using $64^3$ simulations as {\lowres}, which are $\simeq 8$ times cheaper than $128^3$ simulations.
   \textbf{(Green):} using $256^3$ simulations as {\lowres}, which are $\simeq 8$ times most expensive than $128^3$ simulations.
   Shaded area shows the maximum and minimum errors among ten test simulations.}
   \label{fig:comparison_errors_per_bin_3_changelowres}
\end{figure}

We have so far tested multi-fidelity emulators using $128^3$ simulations ({\lowres}) as low-fidelity and $512^3$ simulations ({\highres}) as high-fidelity.
Figure~\ref{fig:comparison_errors_per_bin_3_changelowres} shows non-linear \xyemulator{50}{3}s using different mass resolutions, $64^3$ and $256^3$ simulations, as low-fidelity.

A $64^3$ simulation is $\simeq 512$ times cheaper than a {\highres} but has a smaller maximum $k$ with $\max{(k)} \simeq 3 \hMpc$.
It produces percent level accuracy for $k \leq 1 \hMpc$ and has worst-case errors $< 5\%$ at small scales $k \geq 1 \hMpc$.
A $256^3$ simulation is $\simeq 8$ times cheaper than a {\highres} simulation,
so the computational cost for a \xyemulator{50}{3} is $\simeq 9$ {\highres} simulations.
This emulator mildly outperforms the emulator where {\lowres} is $128^3$,  with an average percent-level emulation until $k \simeq 12 \hMpc$, but at a substantially increased computational cost. % Figure~\ref{fig:single_multi_comparison_3_11_nonlinear} shows it still has substantially improved accuracy than a single fidelity emulator with $9$ {\highres} simulations, which has similar computational cost.
%Since simulators usually test their simulations with multiple resolutions before submitting their biggest job, it is worth testing an emulator using an intermediate resolution as {\lowres}.

% Figure~\ref{fig:comparison_errors_per_bin_3_changelowres} \xyemulator{50}{3} with $256^3$ simulations as low-fidelity is percent-level accurate until $k \simeq 10 \hMpc$ on average.
% For the worst-case error, it shows $\lesssim 0.5\%$ error per $k$ bin at $k \leq 1 \hMpc$ and $\lesssim 1 \%$ at $k \geq 4 \hMpc$ while having a larger error $\lesssim 2\%$ per $k$ bin at intermediate scales, $1 \hMpc < k < 4 \hMpc$.

Figure~\ref{fig:comparison_errors_per_bin_3_changelowres} demonstrates that one can fuse various qualities of {\lowres} with {\highres} simulations to build a multi-fidelity emulator.
Figure~\ref{fig:comparison_errors_per_bin_3_changelowres} also shows that the multi-fidelity emulator's accuracy depends on the correlation between {\lowres} and {\highres}. A $64^3$ simulation is only a rough approximation to its $512^3$ counterpart, so the emulator that uses $64^3$ simulations as low-fidelity is less accurate than the others in Figure~\ref{fig:comparison_errors_per_bin_3_changelowres}.

\subsubsection{Emulation at $z = 1$ and $z =2$}
\label{subsec:higher_z}

% \begin{figure}
%    \includegraphics[width=\columnwidth]{images/comparison_errors_per_bin_3(z0_z1)_lowRes.pdf}
%    \caption{Relative emulator errors for a non-linear emulator at $z = 0$ and $z =1$.
%    Shaded area indicates the maximum and minimum errors.}
%    \label{fig:non_linear_mf_z_1}
% \end{figure}
\begin{figure}
   \includegraphics[width=\columnwidth]{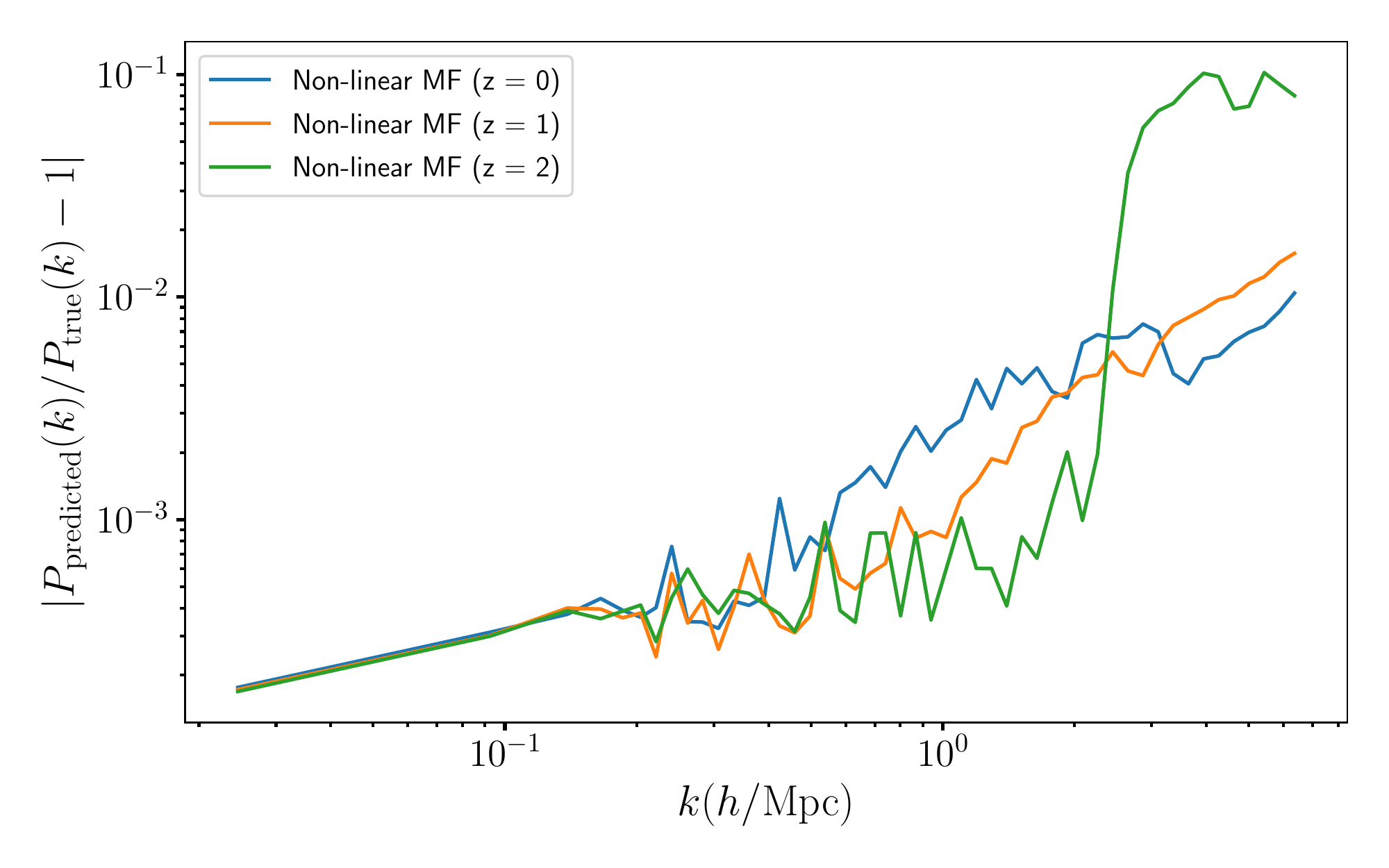}
   \caption{Relative emulator errors for a non-linear emulator at different redshifts, $z \in \{0, 1, 2\}$.
   Note the y-axis is in $\log_{10}$ scale.
   The larger error in the $z = 2$ emulator at $k > 2 \hMpc$ may be due to a transient near the mean-particle spacing in the {\lowres} simulations, see Figure~\ref{fig:particle_spacing_z2}.}
   \label{fig:non_linear_mf_z_0_1_2}
\end{figure}

This section examines the performance of a non-linear emulator at higher redshifts, $z = 1$ and $z = 2$.
Figure~\ref{fig:non_linear_mf_z_0_1_2} shows the emulator error of a non-linear \xyemulator{50}{3} at $z = 0, 1, 2$.
The mean error at $z = 1$ is smaller than the $z = 0$ error at $k \leq 2 \hMpc$ while it is larger for $k > 2 \hMpc$.
This result shows that it is easier to train the correlation between fidelities at large scales $k \leq 2 \hMpc$ while harder to train at small scales $k > 2 \hMpc$.
The emulator at $z = 2$ also shows a better performance than $z =0$ at large scales, $k \leq 2 \hMpc$, but the error diverges to $\sim 10\%$ on smaller scales, $k > 2 \hMpc$.
The improved performance on large scales may be because at higher redshifts the matter power spectrum is closer to linear theory and so the correlation between fidelities is easier to learn.

\begin{figure}
   \includegraphics[width=\columnwidth]{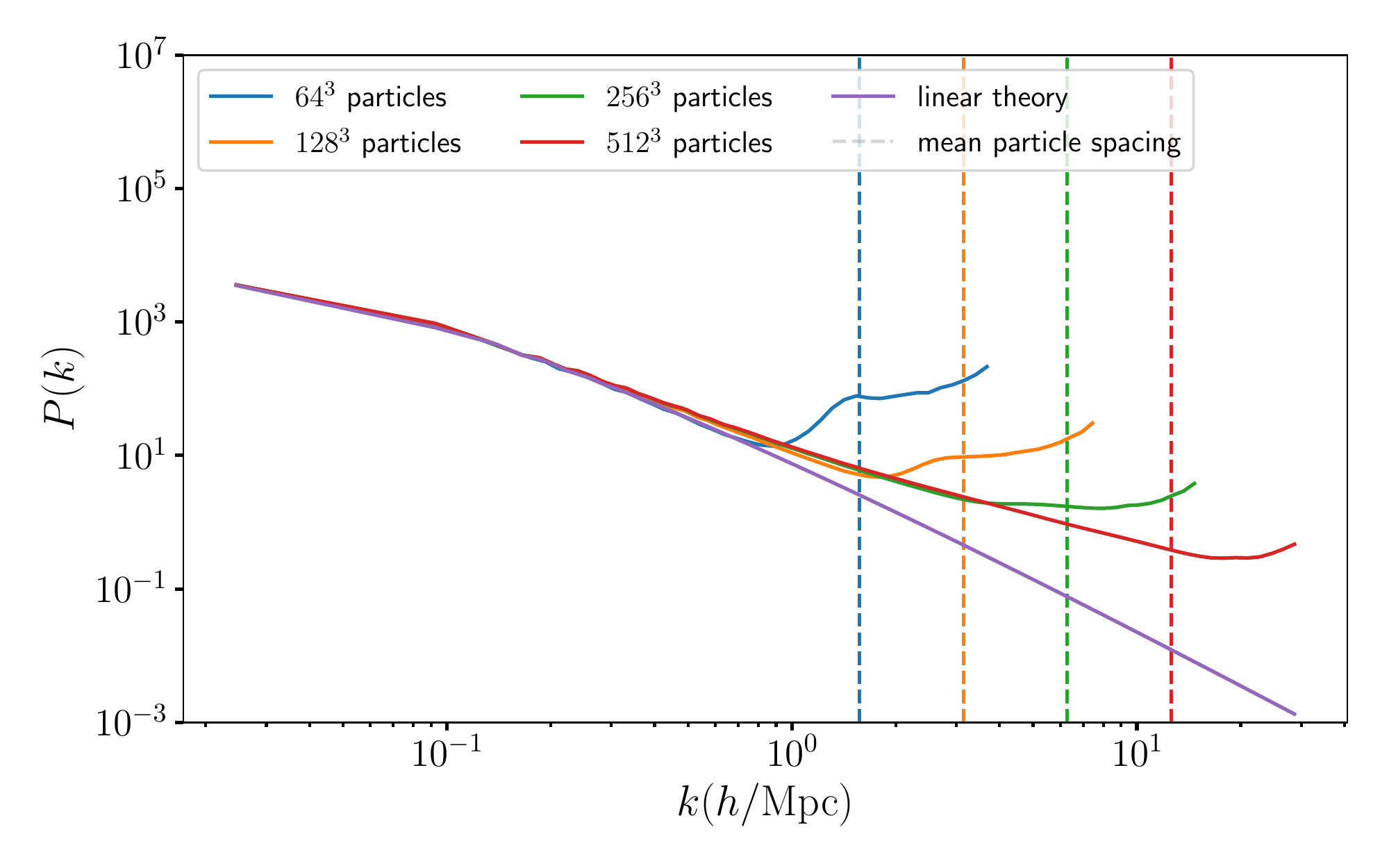}
   \caption{The matter power spectrum at $z = 2$, output by {\mpgadget} with different mass resolutions.
   The vertical dash lines indicated the mean particle spacing $k_\mathrm{spacing}$
   for a given mass resolution.
   \textbf{(Blue):} The matter power spectrum from dark-matter only {\mpgadget}
   simulation with $\npart = 64$.
   \textbf{(Orange):} The matter power spectrum from {\mpgadget} with $\npart = 128$.
   \textbf{(Green):} The matter power spectrum from {\mpgadget} with $\npart = 256$.
   \textbf{(Red):} The matter power spectrum from {\mpgadget} with $\npart = 512$.
   \textbf{(Purple):} Linear theory power spectrum.
   }
   \label{fig:particle_spacing_z2}
\end{figure}

Figure~\ref{fig:particle_spacing_z2} shows the matter power spectrum at $z=2$, with the same cosmological parameters as Figure~\ref{fig:particle_spacing} and indicates a potential explanation. At $z=2$, the low-fidelity simulation contains a systematic at the scale of the mean inter-particle spacing, related to the initial spacing of particles on a regular grid. This systematic is a transient and disappears by $z=0$. However, at redshifts where it is present it implies that the low fidelity simulations contain very little cosmological information on scales near their mean interparticle spacing, $k \simeq 3 \hMpc$ and thus cannot significantly improve the emulation accuracy. It may be possible to improve performance at high redshift with the use of other pre-initial conditions such as a Lagrangian glass \citep{White:1994}.

% However,
% at a higher redshift,
% low-fidelity simulations would have less information at the smallest scales.
% Redshifts closer to the simulation start time ($z = 99$) would have less information beyond the initial particle spacing scale. There is no enough time for particles to evolve and interact with each other.
% Thus, the scales beyond the initial particle spacing are dominated by the artificial initial particle spacing,
% making it harder to extract the correlation between fidelities.
% As shown in Figure~\ref{fig:particle_spacing_z2},
% {\lowres} power spectrum starts to deviate from its {\highres} counterpart from $k \geq 2$.
% Not enough information beyond initial particle spacing is the reason why we observe an improvement at large scales while decreasing accuracy at small scales, as demonstrated in Figure~\ref{fig:non_linear_mf_z_0_1_2}.
% \todo{Simeon, could you confirm my reasoning is correct?} \spb{Yes but I reworded a bit}

% \subsection{Test on additional high-fidelity mean functions}
% \label{subsec:mean_functions}

\section{Runtime}
\label{sec:runtime}

We ran our simulations using {\mpgadget} on UCR's High Performance Computing Center ({\hpcc}) and the Texas Advanced Computing Center (\tacc).
The standard computational setup was $256$ {\mpi} tasks per simulation for both {\highres} ($512^3$ dark matter particles) and {\lowres} ($128^3$ dark matter particles).
The runtime was $\sim 20$ core hours for {\lowres} and $\sim 2\,000$ core hours for {\highres},
with a fixed boxsize $256 \Mpch$.
The computational time for a $64^3$ simulation was $\sim 1.5$ core hours with $64$ {\mpi} tasks and $\sim 280$ core hours for a $256^3$ simulation with $256$ {\mpi} tasks.

The computational cost for training a non-linear \xyemulator{50}{3} (NARGP) was $\simeq 0.5$ hours and $\simeq 1.6$ hours for a linear \xyemulator{50}{3} (AR1) on a single core.
For a single-fidelity emulator, it was $\simeq 2$ minutes on one core.
The compute time could be further improved by parallelizing the hyperparameter optimization for each $k$ bin.
The compute time for optimizing the choice of {\highres} using low-fidelity emulators was $\sim 3$ hours for selecting $3$ {\highres} (on one core).
The run time was $\simeq 12$ seconds for evaluating $10$ test simulations.

% 437 seconds for running 49 low-fidelity emulators on selecting
% two optimal choices for training.
% to run (50 2) combinations, we need ~437*25 seconds = 3 hours (one core),

\section{Conclusions}
\label{sec:conclusions}

We have presented multi-fidelity emulators for the matter power spectrum. Multi-fidelity methods fuse together $N$-body simulations from different mass resolutions to improve interpolation accuracy.
Multi-fidelity emulators use many low-fidelity simulations to learn the power spectrum's dependence on cosmology, correcting for their low resolution by adding a few high-fidelity simulations.
The result is equivalent in accuracy to a single-fidelity emulator performed entirely with much more costly high-fidelity simulations.
A multi-fidelity emulator's physical motivation can be understood using the halo model: low-fidelity simulations capture the two-halo term at large scales, while a few high-fidelity simulations are used to learn the (almost cosmology independent) one-halo term at small scales.

We have also proposed a new sampling strategy which uses low-fidelity simulations as a prior to place high-fidelity training simulations.
We choose our high-fidelity training samples by optimizing the low-fidelity emulator's error.
In this way, the input parameters at which to run {\highres} simulations can be optimized without knowledge of the {\highres} output.
We showed that single-fidelity emulator errors are correlated between different fidelities, indicating that a lower fidelity emulator can serve as a good prior for picking {\highres} simulation points.

Our best multi-fidelity emulator achieved percent level accuracy using only $3$ {\highres} simulations and $50$ {\lowres} simulations, with a total computational cost $\lesssim 4$ {\highres} simulations.
We showed it outperforms a single-fidelity emulator with $11$ {\highres} simulations. We expect that a single-fidelity emulator would require $\sim 50$ {\highres} simulations to compete with the multi-fidelity one at large scales, $k \leq 2 \hMpc$.

In this paper, we used $128^3$ simulations as our low-fidelity training sample and $512^3$ simulations as high-fidelity, with a fixed $256 \Mpch$ box.
However, Figure~\ref{fig:comparison_errors_per_bin_3_changelowres} indicates our method still has a good performance when extended to other resolutions.
We tested our emulator with a series of $10$ {\highres} simulations in a Latin hypercube.
Two types of multi-fidelity emulators, linear (AR1) and non-linear (NARGP), are used.
We showed that both emulators perform similarly at large scales, while the non-linear one has a better accuracy at small scales.

We focussed on $z=0$, but also investigated higher redshifts. Higher redshift power spectra behave more linearly than at $z = 0$, so it is easier to learn the large-scale correlation between fidelities. However, the low-fidelity power spectra are less reliable beyond the mean particle spacing at higher redshifts, inducing some difficulty modelling small scales with $k > 2 \hMpc$.

Our multi-fidelity emulators could provide percent-level predictions for future space- and ground-based surveys at a minimum computational cost.
All current emulators are single-fidelity, training only on expensive high-fidelity simulations.
A single-fidelity emulator requires at least $\sim 40$ simulations to give percent-level accuracy in a $\Lambda$CDM Universe. For example, \cite{Heitmann:2009} use $37$ simulations to emulate a $5$-dimensional $\Lambda$CDM model.
\cite{Euclid:2020} use $\sim 200$ high-fidelity simulations ($3000^3$ dark matter particles) to achieve the upcoming Euclid mission's desired accuracy in an $8$ dimensional parameter space.

% say we can build MF emulator based on existing simulations
Our multi-fidelity methods can also be used to improve the existing single-fidelity emulators.
For example,
suppose we have run $50$ high-resolution simulations to build an emulator.
We can perform $3$ additional super high-resolution simulations and combine them to build a super-resolution multi-fidelity emulator.
The choice of these $3$ simulations could be selected via the optimization strategy proposed in this paper.
Instead of performing super high-resolution simulations,
one could use generative adversarial network techniques \cite[see][]{Li:2020} to generate super-resolution simulations and combine them with a multi-fidelity emulator.

% Combine with current public emulator, using LR simulations
Besides increasing the resolution, multi-fidelity methods could also be used to decrease the emulation uncertainty of an existing emulator by extending it with many low resolution simulations. This indicates a low-cost way to enhance current emulators. Multi-fidelity emulators may make possible efficient expansion of the prior parameter volume. Since high-fidelity simulations are only used to calibrate the resolution, they might not need to span the whole parameter space, implying we can expand the sampling range of an existing emulator by extending the low-fidelity sampling range. We will leave this technique to future work.

% other resolution, other statistics
In this work, we have tested our multi-fidelity emulators with $512^3$ resolution and a relatively small box $256 \Mpch$.
In future we will apply the framework developed here to create a production quality emulator using higher particle load simulations (e.g., $2048^3$ particles) in larger boxes. Other summary statistics, including the halo mass function and the cosmic shear power spectrum, could also be emulated using the same framework.

% astrophysics
The multi-fidelity framework may also be extended to hydrodynamical simulations, which are much more costly than their dark matter-only counterparts.
% As sub-grid physics are resolution dependent, emulation across fidelities might be hard if the grid calibration is not adequate.
%to perform hydrodynamical simulations including gas particles, and it is almost impossible to perform $\sim 200$ high-fidelity hydro-simulations.
% In \cite{Bird:2019}, only $21$ small hydro-simulations ($2 \times 256^3$ particles, $40$ {\Mpch} box) were affordable, even though {\Lya} simulations are relatively cheaper than galaxy formation simulations.
No production hydrodynamical emulators including galaxy formation effects such as AGN feedback yet exist.\footnote{\cite{VillaescusaNavarro:2020} has a neural net emulator trained with $4\,233$ (magneto-)hydrodynamical simulations in a relatively small box, $25 \Mpch$. \cite{Arico:2020} has an hydro-emulator using baryonification methods for BACCO simulations.} However, AGN feedback significantly affects the matter power spectrum at $k > 0.1 \hMpc$ \citep{vanDaalen:2011} and pressure forces can affect the power spectrum at $k \sim 10 \hMpc$ \citep{White:2004}. Thus practical exploitation of the small-scale information from future surveys will require the development of hydrodynamical emulators. By decreasing the computational cost of an emulator by a factor of $\approx 3$ and still outperforming a single-fidelity emulator, the work presented here makes emulation development substantially more practical.

\section*{Software}

We used the \texttt{GPy} \citep{gpy2014} package for Gaussian processes.
For multi-fidelity kernels, we moderately modified the multi-fidelity submodule from \texttt{emukit} \citep{Emukit:2019}.\footnote{\url{https://github.com/EmuKit/emukit}}
We used the {\mpgadget} \citep{MPGADGET:2018} software for simulations.\footnote{\url{https://github.com/MP-Gadget/MP-Gadget}}
We generated customized dark matter-only simulations using Latin hypercubes a modified version of \texttt{SimulationRunner}.\footnote{\url{https://github.com/sbird/SimulationRunner}}

\section*{Data Availability}

The code to reproduce a \xyemulator{50}{3} is available at \url{https://github.com/jibanCat/matter_multi_fidelity_emu} alongside the power spectrum data.

\section*{Acknowledgements}

We thank the referee for providing insightful suggestions and comments.
We thank Martin Fernandez, Phoebe Upton Sanderbeck, Mahdi Qezlou, and Shan-Chang Lin for valuable help and discussions on this project.
We thank Cosmology from Home 2020 for providing a valuable place for discussing simulation-based inference during the pandemic.
MFH acknowledges funding from a NASA FINESST grant.
SB was supported by NSF grant AST-1817256. Computing resources were provided by NSF XSEDE allocation AST180058.

% Reference
\bibliography{sample}

% Appendices are listed in the followings
% \appendix

\label{lastpage}

\end{document}